\documentclass[notitlepage,twocolumn,longbibliography,prl]{revtex4-1}
\usepackage{graphicx}
\graphicspath{{figures/}}
\usepackage{amsfonts}
\usepackage{amssymb}
\usepackage{mathrsfs}
\usepackage{soul}
\usepackage[usenames, dvipsnames]{color}
\usepackage[colorlinks=true,citecolor=blue,urlcolor  = purple]{hyperref}

\usepackage{tikz}

\usepackage[mathcal]{euscript}

\usepackage{amsmath}

\newcommand{\Tr}{\operatorname{Tr}}
\newcommand{\tr}{\operatorname{Tr}}

\newcommand{\ketbra}[2]{| #1 \rangle \langle #2 |}

\newcommand{\abs}[1]{\left| #1\right|}
\newcommand{\mean}[1]{\langle #1 \rangle}
\newcommand{\norm}[1]{\left\| #1 \right\|}

\newcommand{\kommentar}[1]{}

\usepackage{amsthm}

\newtheorem*{lemma*}{Lemma}
\newtheorem*{corollary*}{Corollary}
\theoremstyle{remark}

\newtheorem{example}{Example}

\usepackage{graphicx}

\usepackage{float}
\usepackage{color}
\definecolor{npurple}{rgb}{0.3,0,0.6}

\newcommand{\I}{\openone}

\newcommand{\RR}{\mathbb{R}}

\newcommand{\X}{\mathcal{X}}

\newcommand{\x}{\pmb{x}}

\definecolor{mygray}{gray}{0.6}

\newcommand{\new}[1]{{\color{red} #1}}
\newcommand{\newnew}[1]{{#1}}

\usepackage[utf8]{inputenc}
\usepackage[T1]{fontenc}
\usepackage{multirow}

\usepackage{color}
\usepackage{graphicx}
\usepackage{xcolor}
\definecolor{dark-gray}{rgb}{.35,.55,.55}
\definecolor{dark-blue}{rgb}{.0,.0,.6}

\usepackage[]{hyperref}

\renewcommand{\new}[1]{{ \color{black} #1}}

\usepackage{pdfpages}
\makeatletter
\AtBeginDocument{\let\LS@rot\@undefined}
\makeatother

\begin{document}
\title{Optimising shadow tomography with generalised measurements}
\author{H. Chau Nguyen}
\email{chau.nguyen@uni-siegen.de}

\author{Jan Lennart B\"onsel}
\email{jan.boensel@uni-siegen.de}

\author{Jonathan Steinberg}
\email{steinberg@physik.uni-siegen.de}

\author{Otfried G\"uhne}
\email{otfried.guehne@uni-siegen.de}

\affiliation{Naturwissenschaftlich--Technische Fakult\"{a}t,
Universit\"{a}t Siegen, \\ Walter-Flex-Stra{\ss}e 3, 57068 Siegen, Germany}
\date{\today}

\begin{abstract}
Advances in quantum technology require scalable techniques to efficiently 
extract information from a quantum system. 
Traditional tomography is limited to a handful of qubits 
and shadow tomography has been suggested as a scalable replacement for 
larger systems. Shadow tomography is conventionally analysed based on 
outcomes of ideal projective measurements on the system upon application 
of randomised unitaries. Here, we suggest that shadow tomography can be 
much more straightforwardly formulated for generalised measurements, or 
positive operator valued measures. Based on the idea of the least-square 
estimator shadow tomography with generalised measurements is both more 
general and simpler than the traditional formulation with randomisation 
of unitaries. 
In particular, this formulation allows us to analyse 
theoretical aspects of shadow tomography in detail.  For example, we provide a detailed study of the implication of symmetries in shadow 
tomography. 
Moreover, with this generalisation we also demonstrate how 
the optimisation of measurements for shadow tomography tailored toward a particular set of observables can be carried 
out. 
\end{abstract}

\maketitle

{\it Introduction---} Quantum technology is based on our ability to manipulate quantum mechanical 
states of well-isolated systems: to encode, to process and to extract 
information from the states of the system. Extracting information in this 
context means to design and perform measurements on the system so that 
observables or other properties of the system such as its entropy can be 
inferred. Naively, one may attempt to perform tomography of the state of 
the system.  This amounts to making a sufficiently large number 
of different measurements on the system so that the density operator describing the 
state can be inferred~\cite{smithey_wigner_dist,james_measurement_qubit,heaffner_multiparticle_ions,schwemmer_errors_tomography,paris_quantum_state_estimation}.

However, when considering how the density operator is used later on, 
the entire information contained in the density operator is often not 
needed~\cite{aaronson_journal_2020}.  
In fact, it is impractical to even write down the density 
operator when the number of qubits is large since the dimension of the many-qubit 
system increases exponentially. 
In practice, most often one is not interested in the elements of the 
density operator itself, but rather in certain properties of the quantum 
state, such as the mean values of certain observables or its entropy.
Aiming at inferring directly the observables, bypassing the reconstruction of the 
density operator, shadow tomography has been theoretically 
proposed~\cite{aaronson_journal_2020}.
\citet{huang_predicting_2020} thereafter suggested a practical procedure 
to realise this aim, which has attracted a lot of attention in the 
contemporary research of quantum information processing.

The idea of the protocol is simple. Traditionally, quantum state 
tomography is thought to be only useful once one has accurate 
enough statistics of measurements. However, state estimators such 
as the least square estimator can actually be carried out in 
principle for arbitrary diluted data~\cite{guta_fast_2020}, a fact
well established in data science and machine learning ~\cite{mehta_machine_learning, bishop_machine_learning}.
Indeed, a single data point can contribute a noisy estimate 
of the state; and the final estimated state is obtained by 
averaging over all the data points. Expectedly, when the data 
is diluted, the estimated quantum state can be highly noisy and 
far away from the targeted actual state in the high dimensional state space. This noisy estimation is, 
however, sufficient to predict certain observables or properties 
of the quantum states accurately~\cite{aaronson_journal_2020,huang_predicting_2020}.
Crucially, estimation of observables and certain properties of the 
quantum states for single data points can also be processed without 
explicitly writing down the density operator~\cite{huang_predicting_2020}.
This endows the technique with the promise of scalability.

As for collecting data,~\citet{huang_predicting_2020} suggested to perform random unitaries from a certain chosen set of unitaries on the system and perform a standard ideal measurement afterwards. 
This is equivalent to choosing randomly a measurement from a chosen set.
Since then, various applications of the technique have been found in energy estimation~\cite{hadfield_adaptive_2021,hadfield_measurements_2020}, entanglement detection~\cite{elben_mixed-state_2020,neven_symmetry-resolved_2021}, metrology~\cite{rath_quantum_2021},  analysing scrambled data~\cite{garcia_quantum_2021} and quantum chaos~\cite{joshi_probing_2022}, to name a few. 
Further developments to improve the performance of the scheme ~\cite{huang_efficient_2021,elben_mixed-state_2020,zhang_experimental_measurement_shadows_2021,chen_robust_shadow_estimation_2021,hu_hamiltonian_driven_2022,hu_scrambled_dynamics_2022}  and generalisation to channel shadow tomography have also been proposed~\cite{levy_classical_2021,helsen_estimating_2021}.
In this work, we propose a general framework for shadow tomography with so-called generalised measurements (or POVMs). This theoretical framework contains the randomisation of unitaries as a special case, and at the same time allows for analysis of unavoidable noise in realistic quantum measurements~\cite{arute_quantum_2019,Chen2019}, where projective measurements may not be available.
To our knowledge, there is so far a single proposed procedure for shadow tomography with generalised measurements~\cite{acharya_journal_2022}.
{The suggested procedure is, however, based on application of the original construction of classical shadows in Ref.~\cite{huang_predicting_2020} upon manually synthesising the post-measurement states for generalised measurements.
On the contrary, here we show that classical shadows for generalised measurement can be derived straightforwardly from the least-square estimator~\cite{guta_fast_2020}, which requires no further assumptions on the post-measurement states and contains the framework for ideal measurements  
as a special case~\cite[Appendix A-D]{supp}.} 
In fact, our proposed framework turns out to be much more general and at the same time simpler than randomisation of unitaries.

{\it Shadow tomography with generalised measurements---}
Consider a quantum system of dimension $D$, 
which can either be a single qubit or many qubits. 
A generalised measurement $E$ on the system (positive operator valued measure - POVM) is a collection of positive operators called effects, $E=\{E_1,E_2,\ldots,E_N\}$, summing up to the identity, $\sum_{k=1}^{N} E_k = \I$.
Each generalised measurement $E$ defines a map $\Phi_E$ which maps a density operator $\rho$ to a probability distribution over measurement outcomes,
\begin{equation}
\Phi_E(\rho) = \{\tr(\rho E_k)\}_{k=1}^{N}.
\end{equation}
When the measurement is performed,  
an outcome $k$ is obtained according to this distribution.

\begin{figure}
\centering
\includegraphics[width=0.46\textwidth]{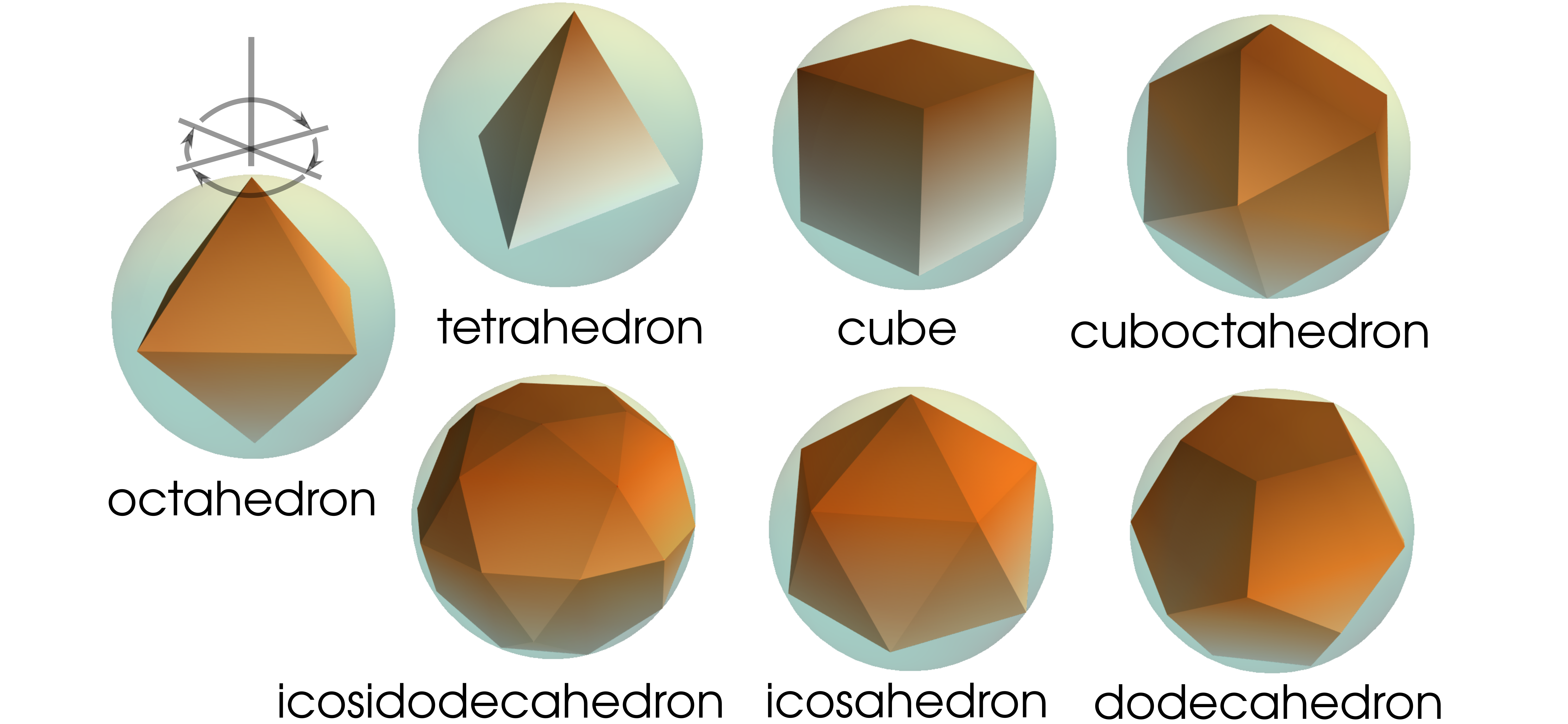}
\caption{Example of generalised measurements defined by polytopes on the Bloch sphere: octahedron ($N=6$), tetrahedron ($N=4$), cube ($N=8$), cuboctahedron ($N=12$), icosahedron ($N=12$), dodecahedron ($N=20$), icosidodecahedron ($N=30$). 
}
\label{fig:platonic}
\end{figure}

Typical measurements in standard quantum mechanics are generalised measurements whose effects $E_k$ are rank-$1$ projections, referred to as \emph{ideal measurements}. 
For example, the measurement of a Pauli operator $\sigma_x$ is an ideal measurement, whose effects are projections on the spin states in the $x$ direction, $\{\ketbra{x^+}{x^+},\ketbra{x^-}{x^-}\}$.
On the other hand, randomising three Pauli measurements $\sigma_x$, $\sigma_y$, $\sigma_z$ is equivalent to a generalised measurement with effects proportional to the projections on the spin states in the $x$, $y$ and $z$ direction, $1/3 \times \{\ketbra{x^\pm}{x^\pm},\ketbra{y^\pm}{y^\pm},\ketbra{z^\pm}{z^\pm}\}$; see~\cite[Appendix~D]{supp} for further discussion.
Since these effects form an octahedron on the Bloch sphere, we also refer to it as the octahedron measurement.  
Generalised measurements, however, allow for much more flexible ways of extracting information from the system.
For example, one can consider the generalised measurements defined by different polytopes as in Fig.~\ref{fig:platonic}.
Generalised measurements become also indispensable in modelling realistic noise in measurement implementation.

Repeating the measurement $M$ times on the system results in a string of outcomes $\{k_i\}_{i=1}^{M}$.
Shadow tomography starts with associating a single outcome $k$ to a distribution $\vec{q}_k= \{\delta_{kl}\}_{l=1}^{N}$ in $\RR^N$.
Such a single data point can be used to obtain a noisy estimate of $\rho$, called \emph{classical shadow}~\cite{aaronson_journal_2020,huang_predicting_2020}
, $\hat{\rho}_k = \chi (\vec{q}_k)$.
As a general strategy in data science, one can require the shadow estimator $\chi$ to be the least-square estimator, of which the solution is well-known~\cite{bishop_machine_learning,guta_fast_2020} (see also \cite[Appendix A]{supp}),
\begin{equation}
\chi_{\text{LS}} = (\Phi_E^{\dagger} \Phi_E)^{-1} \Phi_E^{\dagger}.
\end{equation}
Here we assume that the effects $\{E_k\}_{k=1}^{N}$ span the whole operator space, in which case $E$ is said to be \emph{informationally complete}, so that $\Phi_E^{\dagger} \Phi_E$ is invertible. 
Notice that $\Phi^{\dagger} (\vec{q}_k)=E_k$, and if we define $C_E = \Phi_E^\dagger \Phi_E$, then $C_E (\rho) = \sum_{k=1}^{N} \tr (\rho E_k) E_k$.
The classical shadow can therefore also be written as
\begin{equation}
\hat{\rho}_k = C_{E}^{-1} (E_k).
\label{eq:shadow}
\end{equation}

The classical shadow in Eq.~\eqref{eq:shadow} resembles that in Ref.~\cite{huang_predicting_2020}. 
However, when the measurement is not ideal, the effects $E_k$ do not represent the state of the system after the measurement.
In fact, unlike the procedure proposed in Ref.~\cite{acharya_journal_2022}, our derivation suggests that the states of the system after the measurement is here not important~\cite[Appendix A-C]{supp}.

{\it Estimation of observables and the shadow norm---} 
Each of the classical shadow~\eqref{eq:shadow} serves as an intermediate processed data point for further computation of observables.
Given an observable $X$, each of the classical shadows $\hat{\rho}_k$ gives an estimate for the mean value $\mean{X}$ as  $\hat{x}_k = \tr(\hat{\rho}_{k} X)$.
With the whole dataset $\{k_i\}_{i=1}^{M}$, the average $1/M\sum_{i=1}^{M} \hat{\rho}_{k_i}$ converges to $\rho$~\cite[Appendix A]{supp}, 
therefore $1/M \sum_{i=1}^{M} \hat{x}_{k_i}$ converges to $\mean{X}$.
In this way, the mean value $\mean{X}$ can be estimated. 
For further refinement using the median-of-means estimation and estimation of polynomial functions of the density operator, see Ref.~\cite{huang_predicting_2020}.
As also noted there, the asymptotic rate of convergence of the estimation is related to the variance of the estimator. 
For an observable $X$, the variance of the estimator can be computed as 
$ \mathrm{var} (\hat{x}_k) = \sum_{k=1}^{N} \tr (\hat{\rho}_{k} X)^2 \tr (\rho E_k) - \mean{X}^2$.
Ignoring the second term results in an upper bound for the variance, and finally assuming the worst case scenario, i.e., maximisation over $\rho$, one arrives at the definition of the shadow norm of $X$~\cite{huang_predicting_2020},
\begin{equation}
\mathrm{var} ({\hat{x}_k}) \le \norm{X}_E^2 = \lambda_{\max} \{\sum_{k=1}^{N} \tr (\hat{\rho}_{k} X)^2 E_k \}.
\label{eq:shdnorm}
\end{equation}
where $\lambda_{\max} \{\cdot\}$ denotes the maximal eigenvalue of the corresponding operator. {The estimation procedure applies not only to an observable, but equally well to a set of observables $\X$. Assuming that the observables by certain normalisation all have the same physical unit, the quality of shadow tomography with a generalised measurement $E$ can be characterised by the maximal shadow norm, 
\begin{equation}
    \kappa_E^2 (\X) = \max \{\norm{X}^{2}_E: X \in \X \}.
\end{equation}}
{In the following, we would simply use $\kappa_E^2$ if the set of observables $\X$ is clear.}
{Being the upper bound of the variance of the estimator, the smaller $\kappa_E^2$, the better is the estimator accuracy~\footnote{In practice, the targeted state could be very different from the worst case scenario assumed in obtaining the shadow norm.  
Therefore, it might also be informative to consider the average of the variance with respect to certain ensemble of states.}.} 

{\it Symmetry of generalised measurements and the computation of the classical shadows---}
It has been observed that for certain classes of measurements, the inverse channel $C_E^{-1}$ is particularly simple~\cite{huang_predicting_2020,bu_classical_2022}. 
We are to show that behind this simplicity is the symmetry of the generalised measurement~\cite{nguyen_symmetries_2020,zhu_tomography_2011,bogdanov_statistical_2011}.

{To give the simple intuition, we discuss the example of the octahedron generalised measurement over a qubit plotted in Fig.~\ref{fig:platonic}, leaving the general argument for high dimensional cases in~\cite[Appendix E]{supp}}.
Picking a vertex of the octahedron which corresponds to the effect $E_k$ in Fig.~\ref{fig:platonic}, we consider the symmetry rotations of the octahedron that leave this vertex invariant. 
These are rotations by multiples of $\pi/2$ around the axis going through the chosen vertex. 
Noticeably, there is a single projection (and its complement) that is invariant under these rotations, which corresponds to the state of the spin pointing to the vertex itself. 
In other words, the effect $E_k$ is uniquely specified by the symmetry.
One can show that the corresponding classical shadow $\hat{\rho}_k$ is also invariant under these rotations, which then implies that it is a linear combination of $E_k$ and the identity operator $\I$.
In fact, this is a general property of the so-called uniform and rigidly symmetric measurements defined also for systems of general dimension $D$~\cite{supp,nguyen_symmetries_2020}, {which include in particular the symmetric solids in Fig.~\ref{fig:platonic}}. In all these cases, one has
\begin{equation}
\hat{\rho}_k = a E_k + b \, \openone.
\label{eq:symmetry-inv}
\end{equation} 
The coefficients $a$ and $b$ can be explicitly computed, $a=(D \beta - \alpha^{2})/(D \gamma - \alpha^{3})$ and $b=(\gamma - \alpha \beta) / (D \gamma - \alpha^{3})$, where $\alpha= \tr (E_k)$, $\beta=\tr (E_k^2)$, and $\gamma= \sum_{l=1}^{N} \tr (E_k E_l)^2$ (which are all independent of $k$). 

\begin{figure}
    \centering
    \includegraphics[width=0.46\textwidth]{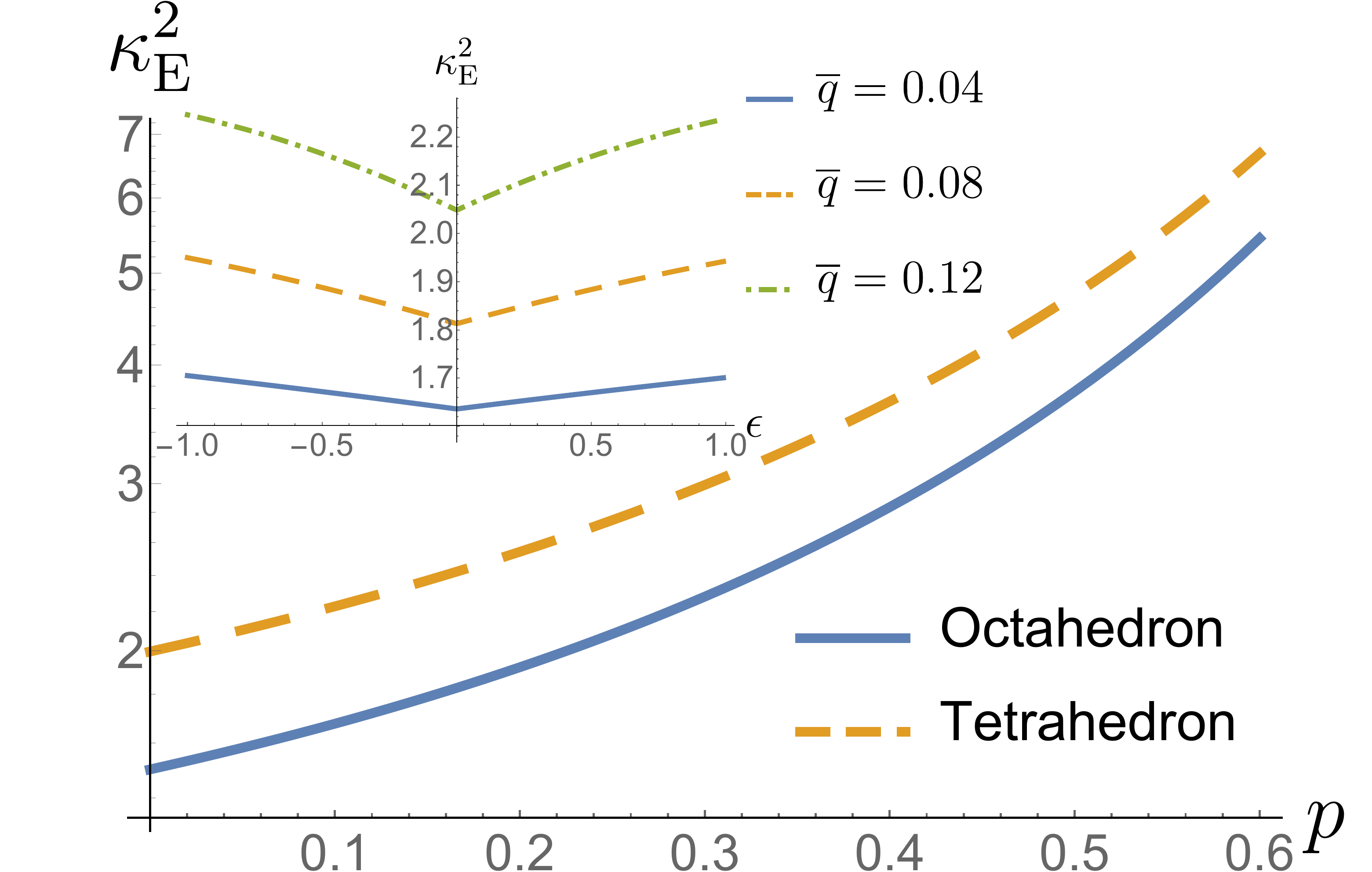}
    \caption{Effects of depolarisation noise (main) and simple readout error noise (inset) on the maximal shadow norm of $128$ pure state projections distributed according to the Haar measure.}
    \label{fig:noise}
\end{figure}

{\it Effects of noise in measurements---}
Measurements in realistic experimental setups aren't ideal. 
The imperfection is due to various sources of noise in setting up the parameters of the measurement devices, or the resolution and the accuracy of readout signals~\cite{arute_quantum_2019,Chen2019}. 
{
As an example, suppose that the measurement $E$ is not perfectly implemented, where the device fails to couple to the system with probability $p$ and indicates an outcome at complete random.
This can be modelled by the effects that are depolarised as 
$\{p \I/N  + (1-p) E_k\}_{k=1}^{N}$.
Another example is the readout error, which is particularly important for superconducting qubits~\cite{error_multiqubits_2021,active_error_2022}. In a simplified model, an outcome $0$ in the computational basis is misread as $1$ with probability $q_+$, and $1$ misread as $0$ with also probability $q_-$. The error rate averaged over the two bases is $\bar{q}=(q_+ + q_-)/2$ and the asymmetry between them is characterised by $\epsilon = (q_+ - q_-)/(q_+ + q_-)$. The measurement effects of the octahedron measurement implemented by randomisation under this noise becomes $1/3 \{(1-q_\pm) \ketbra{t^\pm}{t^\pm} + q_\mp \ketbra{t^\mp}{t^\mp},t=x,y,z\}$. For further discussion, see~\cite[Appendix  F]{supp}.

\newnew{Our formalism directly takes measurement error correction into account, once the noisy effects with an appropriate model are used instead of the ideal ones. 
To access the quality of the shadow tomography after error correction,} we choose $\abs{\X}=128$ pure state projections distributed according the Haar measure as observables.
The dependence of the maximal shadow norm $\kappa_E^2$ on the noise parameters for the tetrahedron and the octahedron measurements is shown in  Fig.~\ref{fig:noise}.
It is interesting to see that in either cases, the maximal shadow norm $\kappa_E^2$ depends only weakly on small error rate, showing the robustness of shadow tomography against noise. 
}

{\it Optimisation of generalised measurement for shadow tomography---}
{Given a set of observables $\X$, one would like to find the generalised measurement $E$ so that the maximal shadow norm is minimised,
\begin{equation}
E^{\ast} = \arg \min_E \kappa_E^2 (\X).
\label{eq:optim}
\end{equation}}
Restricted to randomised unitaries, this optimisation is impractical to carry out~\cite[Appendix B]{supp}.
Extending to all generalised measurements, this is simply an optimisation over a convex domain. 
We implemented simulated annealing for the minimisation and found the obtained optima to be highly reliable~\cite[Appendix~G]{supp}.
{Below we start with discussing the case of a single qubit, which, despite being simple, is also the basis to understand the case of many qubits.}

\begin{figure}[!hbt]
\centering
\includegraphics[width=0.5\textwidth]{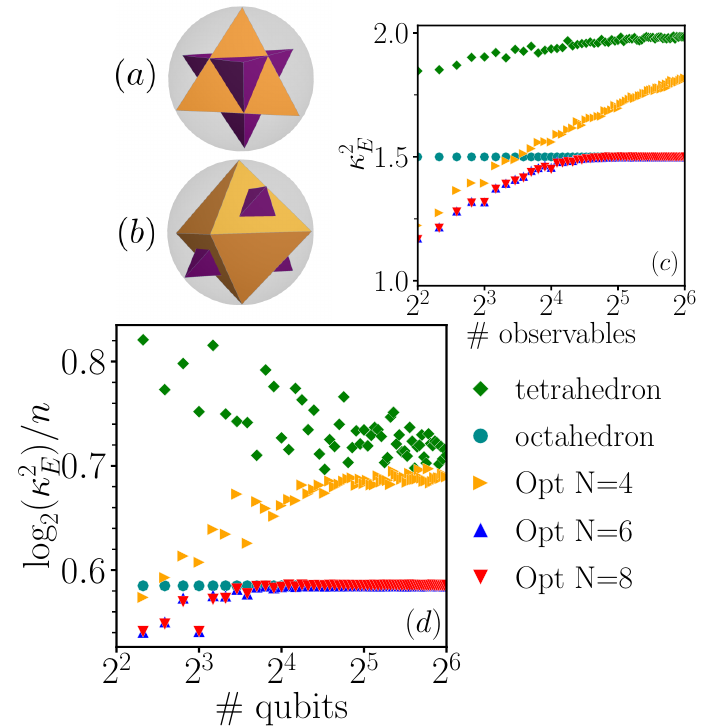}
\caption{Targeted observables and optimal generalised measurements. 
(a) For observables corresponding to four projections defined by the orange tetrahedron, the measurement corresponding to the inverted tetrahedron measurement (violet) is optimal. 
(b) For observables corresponding to eigenprojections of the Pauli observables $\sigma_x$, $\sigma_y$ and $\sigma_z$ (orange octahedron), the violet tetrahedron measurement is optimal.
(c) Optimal shadow norms given by the optimiser (labelled \texttt{Opt} with the number of measurement outcomes) as a function of the number of single-qubit projection observables randomly distributed according to the Haar measure. 
(d) Similarly, optimal shadow norms given by the optimiser as a function of the number of qubits. The observables are tensor products of single-qubit projections distributed according to the Haar measure. In (c) and (d), the shadow norm for the tetrahedron and octahedron measurements are also shown.}
\label{fig:sic-pauli}
\end{figure}

\begin{example}
\label{ex:single-qubit-opt}
Consider a single qubit. {As for the observables $\X$, we consider the following possibilities:}
(a) Take observables to be $4$ projections corresponding to the orange tetrahedron in Fig.~\ref{fig:sic-pauli}a.
The squared shadow norm $\kappa_E^2$ is $2$ for the tetrahedron generalised measurement defined exactly by these $4$ projections, and $3/2$ for the octahedron measurement.
The optimiser suggests that the tetrahedron measurement plotted in violet in Fig.~\ref{fig:sic-pauli}a, obtained by centrally inverting the orange tetrahedron, is optimal with $\kappa_E^2= 1$. 
(b) As observables consider the projections onto the eigenstates of the Pauli observables, see the orange octahedron in Fig.~\ref{fig:sic-pauli}b. 
The octahedron measurement itself gives $\kappa_E^2 = 3/2$. 
Interestingly, the optimiser shows that $\kappa_E^2 = 3/2$ can also be obtained with the tetrahedron generalised measurement of $4$ outcomes  indicated in violet in Fig.~\ref{fig:sic-pauli}b.
(c) Lastly, as observables we consider random projections distributed according to the Haar measure on the Bloch sphere. 
Fig.~\ref{fig:sic-pauli}c presents the shadow norms obtained by the optimiser with respect to the number of observables.
For small number of observables ($\abs{\X} \lesssim 15$), the optimiser always finds measurements with a given number of outcomes significantly better than the standard tetrahedron ($N=4$) or the octahedron measurements $(N=6)$. 
It is interesting to see that if the number of outcome is fixed to be $6$ or $8$, the $\kappa_E^2$ converges to the octahedron measurement with the value of $3/2$. 
\end{example}

\new{The last example (c) hints that the octahedron measurement is somewhat special. Indeed, it turns out the squared shadow norm with respect to the octahedron measurement for any projection is identically $3/2$. Using this fact, we show that if the targeted observables are \emph{all} the projections on arbitrary pure states of the qubit, the optimal measurement would be the octahedron measurement assuming equal trace of the effects~\cite[Appendix H]{supp}.}

{\it Tensoring the shadow construction for many-body systems---}
Shadow tomography is especially designed for the cases where the system size is large. 
Consider the case where the system consists of $n$ qubits, corresponding to the total dimension of $D=2^{n}$.
In this case, shadow tomography can be performed by making (identical or not identical) generalised measurements $\{E^{(1)}, E^{(2)}, \ldots, E^{(n)}\}$ on each of the qubit, each described by a collection of $N_i$ effects, $E^{(i)} = \{E^{(i)}_k\}_{k=1}^{N_i}$.
Theoretically, this corresponds to a measurement of a generalised measurement $E^{\text{tot}}$ on the whole system with each effect labelled by a string of outcomes $\pmb{k}= \{k^{(1)},k^{(2)},\ldots,k^{(n)}\}$,
$E^{\text{tot}}_{\pmb{k}} = 
E^{(1)}_{k^{(1)}} \otimes  
E^{(2)}_{k^{(2)}} \otimes 
\cdots
\otimes
E^{(n)}_{k^{(n)}}$. 
The whole general analysis above can be applied. 
In fact, such a string $\pmb{k}$ of outcomes corresponds simply to the classical shadow
\begin{equation}
\hat{\rho}^{\text{tot}}_{\pmb{k}} = 
\hat{\rho}_{k^{(1)}}^{(1)}
\otimes
\hat{\rho}_{k^{(2)}}^{(2)}
\otimes
\cdots
\otimes
\hat{\rho}_{k^{(n)}}^{(n)},
\end{equation}
where $\hat{\rho}^{(i)}_{k^{(i)}}$  being the classical shadow corresponding to the measurement $E^{(i)}$ on the $i$-th qubit. 
Crucially, the typical observables of the system can be easily estimated without (impractically) explicitly computing the classical shadows in the form of a $D \times D$ matrix~\cite{huang_predicting_2020}.
Indeed, an observable $X$ on the system is often of the form 
$X= X^{(1)} \otimes X^{(2)} \otimes \cdots \otimes X^{(n)}$.
Then, a single string of outcomes $\pmb{k}$ gives rise to a single estimate of $\mean{X}$ as
$\Tr [ \hat{\rho}^{(1)}_{k^{(1)}} X^{(1)}] \,
\Tr [ \hat{\rho}^{(2)}_{k^{(2)}} X^{(2)}] \, 
\cdots
\Tr [ \hat{\rho}^{(n)}_{k^{(n)}} X^{(n)}]$.
The final estimate of  $\mean{X}$ is as usual obtained by averaging over all data points. 
Observe that it is not necessary to construct the large density operator of the whole system.
Moreover, the shadow norm of such a factorised observable also factorises 
$\norm{X}_{E}  = 
\norm{X^{(1)}}_{E^{(1)}} 
\norm{X^{(2)}}_{E^{(2)}} 
\cdots
\norm{X^{(n)}}_{E^{(n)}}$.

{\it Optimising generalised measurements for many-body systems---}
For many qubits, the number of parameters to be optimised in Eq.~\eqref{eq:optim} increases exponentially. 
To simplify, one can assume that for a many-qubit system, the generalised measurement is factorised as a tensor product over the qubits as we discussed above.
Moreover, if there is no preference among the qubits, one can also assume that, $E^{(1)}=E^{(2)} = \cdots = E^{(n)}$.
The complexity of the computation under these assumptions is only linear in the number of qubits and the number of observables. 
\begin{example}
{We consider a system of upto $n=64$ qubits. 
We choose  $\abs{X}=n$ observables which are products of different component observables on single qubits. 
The component observables on single qubits are randomly distributed according to the Haar measure. 
As the qubits are equivalent, one might anticipate that the optimal factorising measurement for the qubits is similar to those that are optimised separately for each qubit.
Our simulations confirm this expectation.
In Fig.~\ref{fig:sic-pauli}d, for small number of qubits ($n \lesssim 10$), the optimiser with $N=6$ and $N=8$ gives significantly lower shadow norms for the choice of tetrahedron or octahedron measurements.
On the other hand, observe that as the number of qubits increases, the obtained optimal shadow norm converges to that given by the octahedron measurement, pointing to the speciality of the octahedron measurement on qubit-based platforms as we discussed in Example~\ref{ex:single-qubit-opt} (c).} 
\end{example}

{\it Conclusion---}
{Being both more general and simpler, the formulation of shadow tomography with generalised measurements sheds light on various aspects of shadow tomography.} 
This also opens a range of interesting questions for future research. 
Further analysis of realistic noise in the existing and future experiments~\cite{fischer_ancilla_free_2022,stricker_2022} of shadow tomography could be considered.
Extension of this framework to channel tomography is of direct interest. 
It would be also important to see whether the technique of derandomisation~\cite{huang_efficient_2021} can also be incorporated.
\new{The optimality of the octahedron measurement for shadow tomography for qubit-based system suggests a connection between geometry and shadow tomography. Investigation of this connection and extension for higher dimensional systems would be an interesting direction.
Also the construction of optimal measurements for nonlinear functions of the density operator, or shadow tomography of a specific set of density operators, is in demand for further applications of shadow tomography.}

\begin{acknowledgments}

{\it Note added:} While finishing this work, we learned that related 
results to estimate non-commuting observables from a special 
generalized measurement has been derived in \cite{mcnulty_2022}. 
After submission of our work to the arxiv, a similar protocol
for shadow tomography using the special example of a SIC POVM has 
been suggested and experimentally implemented in \cite{stricker_2022}. {Both works do not, however, develop a framework for shadow tomography
with arbitrary generalised measurements.
}

The authors would like to thank Kiara Hansenne, Satoya Imai, Matthias Kleinmann, Martin Kliesch (with indirect comments), Micha{\l}   Oszmaniec, Salwa Shaglel, Lina Vandré, Zhen-Peng Xu, and Benjamin Yadin for inspiring discussions and comments. 
The University of Siegen is kindly acknowledged for enabling our computations through the \texttt{OMNI} cluster. 
This work was supported by the Deutsche Forschungsgemeinschaft (DFG, German Research Foundation, project numbers 447948357 and 440958198), the Sino-German Center for Research Promotion (Project M-0294), 
the ERC (Consolidator Grant 683107/TempoQ),
and the German Ministry of Education and Research (Project QuKuK, BMBF Grant No. 16KIS1618K). 
JLB and JS acknowledge support from the House of Young Talents of the University of Siegen.
\end{acknowledgments}

\bibliography{stomography}

\begin{thebibliography}{45}%
\makeatletter
\providecommand \@ifxundefined [1]{%
 \@ifx{#1\undefined}
}%
\providecommand \@ifnum [1]{%
 \ifnum #1\expandafter \@firstoftwo
 \else \expandafter \@secondoftwo
 \fi
}%
\providecommand \@ifx [1]{%
 \ifx #1\expandafter \@firstoftwo
 \else \expandafter \@secondoftwo
 \fi
}%
\providecommand \natexlab [1]{#1}%
\providecommand \enquote  [1]{``#1''}%
\providecommand \bibnamefont  [1]{#1}%
\providecommand \bibfnamefont [1]{#1}%
\providecommand \citenamefont [1]{#1}%
\providecommand \href@noop [0]{\@secondoftwo}%
\providecommand \href [0]{\begingroup \@sanitize@url \@href}%
\providecommand \@href[1]{\@@startlink{#1}\@@href}%
\providecommand \@@href[1]{\endgroup#1\@@endlink}%
\providecommand \@sanitize@url [0]{\catcode `\\12\catcode `\$12\catcode
  `\&12\catcode `\#12\catcode `\^12\catcode `\_12\catcode `\%12\relax}%
\providecommand \@@startlink[1]{}%
\providecommand \@@endlink[0]{}%
\providecommand \url  [0]{\begingroup\@sanitize@url \@url }%
\providecommand \@url [1]{\endgroup\@href {#1}{\urlprefix }}%
\providecommand \urlprefix  [0]{URL }%
\providecommand \Eprint [0]{\href }%
\providecommand \doibase [0]{http://dx.doi.org/}%
\providecommand \selectlanguage [0]{\@gobble}%
\providecommand \bibinfo  [0]{\@secondoftwo}%
\providecommand \bibfield  [0]{\@secondoftwo}%
\providecommand \translation [1]{[#1]}%
\providecommand \BibitemOpen [0]{}%
\providecommand \bibitemStop [0]{}%
\providecommand \bibitemNoStop [0]{.\EOS\space}%
\providecommand \EOS [0]{\spacefactor3000\relax}%
\providecommand \BibitemShut  [1]{\csname bibitem#1\endcsname}%
\let\auto@bib@innerbib\@empty
\bibitem [{\citenamefont {Smithey}\ \emph {et~al.}(1993)\citenamefont
  {Smithey}, \citenamefont {Beck}, \citenamefont {Raymer},\ and\ \citenamefont
  {Faridani}}]{smithey_wigner_dist}%
  \BibitemOpen
  \bibfield  {author} {\bibinfo {author} {\bibfnamefont {D.~T.}\ \bibnamefont
  {Smithey}}, \bibinfo {author} {\bibfnamefont {M.}~\bibnamefont {Beck}},
  \bibinfo {author} {\bibfnamefont {M.~G.}\ \bibnamefont {Raymer}}, \ and\
  \bibinfo {author} {\bibfnamefont {A.}~\bibnamefont {Faridani}},\ }\bibfield
  {title} {\enquote {\bibinfo {title} {Measurement of the {Wigner} distribution
  and the density matrix of a light mode using optical homodyne tomography:
  Application to squeezed states and the vacuum},}\ }\href
  {https://doi.org/10.1103/PhysRevLett.70.1244} {\bibfield  {journal} {\bibinfo
   {journal} {Phys. Rev. Lett.}\ }\textbf {\bibinfo {volume} {70}},\ \bibinfo
  {pages} {1244} (\bibinfo {year} {1993})}\BibitemShut {NoStop}%
\bibitem [{\citenamefont {James}\ \emph {et~al.}(2001)\citenamefont {James},
  \citenamefont {Kwiat}, \citenamefont {Munro},\ and\ \citenamefont
  {White}}]{james_measurement_qubit}%
  \BibitemOpen
  \bibfield  {author} {\bibinfo {author} {\bibfnamefont {D.~F.~V.}\
  \bibnamefont {James}}, \bibinfo {author} {\bibfnamefont {P.~G.}\ \bibnamefont
  {Kwiat}}, \bibinfo {author} {\bibfnamefont {W.~J.}\ \bibnamefont {Munro}}, \
  and\ \bibinfo {author} {\bibfnamefont {A.~G.}\ \bibnamefont {White}},\
  }\bibfield  {title} {\enquote {\bibinfo {title} {Measurement of qubits},}\
  }\href {https://doi.org/10.1103/PhysRevA.64.052312} {\bibfield  {journal}
  {\bibinfo  {journal} {Phys. Rev. A}\ }\textbf {\bibinfo {volume} {64}},\
  \bibinfo {pages} {052312} (\bibinfo {year} {2001})}\BibitemShut {NoStop}%
\bibitem [{\citenamefont {Häffner}\ \emph {et~al.}(2005)\citenamefont
  {Häffner}, \citenamefont {Hänsel}, \citenamefont {Roos}, \citenamefont
  {Benhelm}, \citenamefont {al~kar}, \citenamefont {Chwalla}, \citenamefont
  {Körber}, \citenamefont {Rapol}, \citenamefont {Riebe}, \citenamefont
  {Schmidt}, \citenamefont {Becher}, \citenamefont {Gühne}, \citenamefont
  {Dür},\ and\ \citenamefont {Blatt}}]{heaffner_multiparticle_ions}%
  \BibitemOpen
  \bibfield  {author} {\bibinfo {author} {\bibfnamefont {H.}~\bibnamefont
  {Häffner}}, \bibinfo {author} {\bibfnamefont {W.}~\bibnamefont {Hänsel}},
  \bibinfo {author} {\bibfnamefont {C.~F.}\ \bibnamefont {Roos}}, \bibinfo
  {author} {\bibfnamefont {J.}~\bibnamefont {Benhelm}}, \bibinfo {author}
  {\bibfnamefont {D.~Chek}\ \bibnamefont {al~kar}}, \bibinfo {author}
  {\bibfnamefont {M.}~\bibnamefont {Chwalla}}, \bibinfo {author} {\bibfnamefont
  {T.}~\bibnamefont {Körber}}, \bibinfo {author} {\bibfnamefont {U.~D.}\
  \bibnamefont {Rapol}}, \bibinfo {author} {\bibfnamefont {M.}~\bibnamefont
  {Riebe}}, \bibinfo {author} {\bibfnamefont {P.~O.}\ \bibnamefont {Schmidt}},
  \bibinfo {author} {\bibfnamefont {C.}~\bibnamefont {Becher}}, \bibinfo
  {author} {\bibfnamefont {O.}~\bibnamefont {Gühne}}, \bibinfo {author}
  {\bibfnamefont {W.}~\bibnamefont {Dür}}, \ and\ \bibinfo {author}
  {\bibfnamefont {R.}~\bibnamefont {Blatt}},\ }\bibfield  {title} {\enquote
  {\bibinfo {title} {Scalable multiparticle entanglement of trapped ions},}\
  }\href {https://doi.org/10.1038/nature04279} {\bibfield  {journal} {\bibinfo
  {journal} {Nature}\ }\textbf {\bibinfo {volume} {438}},\ \bibinfo {pages}
  {643–646} (\bibinfo {year} {2005})}\BibitemShut {NoStop}%
\bibitem [{\citenamefont {Schwemmer}\ \emph {et~al.}(2015)\citenamefont
  {Schwemmer}, \citenamefont {Knips}, \citenamefont {Richart}, \citenamefont
  {Weinfurter}, \citenamefont {Moroder}, \citenamefont {Kleinmann},\ and\
  \citenamefont {Gühne}}]{schwemmer_errors_tomography}%
  \BibitemOpen
  \bibfield  {author} {\bibinfo {author} {\bibfnamefont {C.}~\bibnamefont
  {Schwemmer}}, \bibinfo {author} {\bibfnamefont {L.}~\bibnamefont {Knips}},
  \bibinfo {author} {\bibfnamefont {D.}~\bibnamefont {Richart}}, \bibinfo
  {author} {\bibfnamefont {H.}~\bibnamefont {Weinfurter}}, \bibinfo {author}
  {\bibfnamefont {T.}~\bibnamefont {Moroder}}, \bibinfo {author} {\bibfnamefont
  {M.}~\bibnamefont {Kleinmann}}, \ and\ \bibinfo {author} {\bibfnamefont
  {O.}~\bibnamefont {Gühne}},\ }\bibfield  {title} {\enquote {\bibinfo {title}
  {Systematic errors in current quantum state tomography tools},}\ }\href
  {https://doi.org/10.1103/PhysRevLett.114.080403} {\bibfield  {journal}
  {\bibinfo  {journal} {Phys. Rev. Lett.}\ }\textbf {\bibinfo {volume} {114}},\
  \bibinfo {pages} {080403} (\bibinfo {year} {2015})}\BibitemShut {NoStop}%
\bibitem [{\citenamefont {Paris}\ and\ \citenamefont {{\v
  R}eh{\'a}{\v{c}}ek}(2004)}]{paris_quantum_state_estimation}%
  \BibitemOpen
  \bibfield  {author} {\bibinfo {author} {\bibfnamefont {M.}~\bibnamefont
  {Paris}}\ and\ \bibinfo {author} {\bibfnamefont {J.}~\bibnamefont {{\v
  R}eh{\'a}{\v{c}}ek}},\ }\href {https://doi.org/10.1007/b98673} {\emph
  {\bibinfo {title} {Quantum State Estimation}}}\ (\bibinfo  {publisher}
  {Springer, Berlin, Heidelberg},\ \bibinfo {year} {2004})\BibitemShut
  {NoStop}%
\bibitem [{\citenamefont {Aaronson}(2020)}]{aaronson_journal_2020}%
  \BibitemOpen
  \bibfield  {author} {\bibinfo {author} {\bibfnamefont {S.}~\bibnamefont
  {Aaronson}},\ }\bibfield  {title} {\enquote {\bibinfo {title} {Shadow
  tomography of quantum states},}\ }\href
  {https://epubs.siam.org/doi/10.1137/18M120275X} {\bibfield  {journal}
  {\bibinfo  {journal} {SIAM J. Comput.}\ }\textbf {\bibinfo {volume} {49}},\
  \bibinfo {pages} {368--394} (\bibinfo {year} {2020})}\BibitemShut {NoStop}%
\bibitem [{\citenamefont {Huang}\ \emph {et~al.}(2020)\citenamefont {Huang},
  \citenamefont {Kueng},\ and\ \citenamefont
  {Preskill}}]{huang_predicting_2020}%
  \BibitemOpen
  \bibfield  {author} {\bibinfo {author} {\bibfnamefont {H.~Y.}\ \bibnamefont
  {Huang}}, \bibinfo {author} {\bibfnamefont {R.}~\bibnamefont {Kueng}}, \ and\
  \bibinfo {author} {\bibfnamefont {J.}~\bibnamefont {Preskill}},\ }\bibfield
  {title} {\enquote {\bibinfo {title} {Predicting many properties of a quantum
  system from very few measurements},}\ }\href
  {http://www.nature.com/articles/s41567-020-0932-7} {\bibfield  {journal}
  {\bibinfo  {journal} {Nat. Phys.}\ }\textbf {\bibinfo {volume} {16}},\
  \bibinfo {pages} {1050--1057} (\bibinfo {year} {2020})}\BibitemShut {NoStop}%
\bibitem [{\citenamefont {Gu\c{t}\u{a}}\ \emph {et~al.}(2020)\citenamefont
  {Gu\c{t}\u{a}}, \citenamefont {Kahn}, \citenamefont {Kueng},\ and\
  \citenamefont {Tropp}}]{guta_fast_2020}%
  \BibitemOpen
  \bibfield  {author} {\bibinfo {author} {\bibfnamefont {M.}~\bibnamefont
  {Gu\c{t}\u{a}}}, \bibinfo {author} {\bibfnamefont {J.}~\bibnamefont {Kahn}},
  \bibinfo {author} {\bibfnamefont {R.}~\bibnamefont {Kueng}}, \ and\ \bibinfo
  {author} {\bibfnamefont {J.~A.}\ \bibnamefont {Tropp}},\ }\bibfield  {title}
  {\enquote {\bibinfo {title} {Fast state tomography with optimal error
  bounds},}\ }\href {https://doi.org/10.1088/1751-8121/ab8111} {\bibfield
  {journal} {\bibinfo  {journal} {J. Phys. A: Math. Theor.}\ }\textbf {\bibinfo
  {volume} {53}},\ \bibinfo {pages} {204001} (\bibinfo {year}
  {2020})}\BibitemShut {NoStop}%
\bibitem [{\citenamefont {Mehta}\ \emph {et~al.}(2019)\citenamefont {Mehta},
  \citenamefont {Bukov}, \citenamefont {Wang}, \citenamefont {Day},
  \citenamefont {Richardson}, \citenamefont {Fisher},\ and\ \citenamefont
  {Schwab}}]{mehta_machine_learning}%
  \BibitemOpen
  \bibfield  {author} {\bibinfo {author} {\bibfnamefont {P.}~\bibnamefont
  {Mehta}}, \bibinfo {author} {\bibfnamefont {M.}~\bibnamefont {Bukov}},
  \bibinfo {author} {\bibfnamefont {C.-H.}\ \bibnamefont {Wang}}, \bibinfo
  {author} {\bibfnamefont {A.~G.~R.}\ \bibnamefont {Day}}, \bibinfo {author}
  {\bibfnamefont {C.}~\bibnamefont {Richardson}}, \bibinfo {author}
  {\bibfnamefont {C.~K.}\ \bibnamefont {Fisher}}, \ and\ \bibinfo {author}
  {\bibfnamefont {D.~J.}\ \bibnamefont {Schwab}},\ }\bibfield  {title}
  {\enquote {\bibinfo {title} {A high-bias, low-variance introduction to
  machine learning for physicists},}\ }\href
  {https://doi.org/10.1016/j.physrep.2019.03.001} {\bibfield  {journal}
  {\bibinfo  {journal} {Phys. Rep.}\ }\textbf {\bibinfo {volume} {810}},\
  \bibinfo {pages} {1--124} (\bibinfo {year} {2019})}\BibitemShut {NoStop}%
\bibitem [{\citenamefont {Bishop}(2006)}]{bishop_machine_learning}%
  \BibitemOpen
  \bibfield  {author} {\bibinfo {author} {\bibfnamefont {C.~M.}\ \bibnamefont
  {Bishop}},\ }\href {https://link.springer.com/book/9780387310732} {\emph
  {\bibinfo {title} {Pattern Recognition and Machine Learning}}}\ (\bibinfo
  {publisher} {Springer New York},\ \bibinfo {year} {2006})\BibitemShut
  {NoStop}%
\bibitem [{\citenamefont {Hadfield}(2021)}]{hadfield_adaptive_2021}%
  \BibitemOpen
  \bibfield  {author} {\bibinfo {author} {\bibfnamefont {C.}~\bibnamefont
  {Hadfield}},\ }\bibfield  {title} {\enquote {\bibinfo {title} {Adaptive
  {Pauli} shadows for energy estimation},}\ }\href
  {http://arxiv.org/abs/2105.12207} {\bibfield  {journal} {\bibinfo  {journal}
  {arXiv:2105.12207}\ } (\bibinfo {year} {2021})}\BibitemShut {NoStop}%
\bibitem [{\citenamefont {Hadfield}\ \emph {et~al.}(2020)\citenamefont
  {Hadfield}, \citenamefont {Bravyi}, \citenamefont {Raymond},\ and\
  \citenamefont {Mezzacapo}}]{hadfield_measurements_2020}%
  \BibitemOpen
  \bibfield  {author} {\bibinfo {author} {\bibfnamefont {C.}~\bibnamefont
  {Hadfield}}, \bibinfo {author} {\bibfnamefont {S.}~\bibnamefont {Bravyi}},
  \bibinfo {author} {\bibfnamefont {R.}~\bibnamefont {Raymond}}, \ and\
  \bibinfo {author} {\bibfnamefont {A.}~\bibnamefont {Mezzacapo}},\ }\bibfield
  {title} {\enquote {\bibinfo {title} {Measurements of quantum {Hamiltonians}
  with locally-biased classical shadows},}\ }\href
  {http://arxiv.org/abs/2006.15788} {\bibfield  {journal} {\bibinfo  {journal}
  {arXiv:2006.15788}\ } (\bibinfo {year} {2020})}\BibitemShut {NoStop}%
\bibitem [{\citenamefont {Elben}\ \emph {et~al.}(2020)\citenamefont {Elben},
  \citenamefont {Kueng}, \citenamefont {Huang}, \citenamefont {van Bijnen},
  \citenamefont {Kokail}, \citenamefont {Dalmonte}, \citenamefont {Calabrese},
  \citenamefont {Kraus}, \citenamefont {Preskill}, \citenamefont {Zoller},\
  and\ \citenamefont {Vermersch}}]{elben_mixed-state_2020}%
  \BibitemOpen
  \bibfield  {author} {\bibinfo {author} {\bibfnamefont {A.}~\bibnamefont
  {Elben}}, \bibinfo {author} {\bibfnamefont {R.}~\bibnamefont {Kueng}},
  \bibinfo {author} {\bibfnamefont {H.~Y.}\ \bibnamefont {Huang}}, \bibinfo
  {author} {\bibfnamefont {R.}~\bibnamefont {van Bijnen}}, \bibinfo {author}
  {\bibfnamefont {C.}~\bibnamefont {Kokail}}, \bibinfo {author} {\bibfnamefont
  {M.}~\bibnamefont {Dalmonte}}, \bibinfo {author} {\bibfnamefont
  {P.}~\bibnamefont {Calabrese}}, \bibinfo {author} {\bibfnamefont
  {B.}~\bibnamefont {Kraus}}, \bibinfo {author} {\bibfnamefont
  {J.}~\bibnamefont {Preskill}}, \bibinfo {author} {\bibfnamefont
  {P.}~\bibnamefont {Zoller}}, \ and\ \bibinfo {author} {\bibfnamefont
  {B.}~\bibnamefont {Vermersch}},\ }\bibfield  {title} {\enquote {\bibinfo
  {title} {Mixed-state entanglement from local randomized measurements},}\
  }\href {https://link.aps.org/doi/10.1103/PhysRevLett.125.200501} {\bibfield
  {journal} {\bibinfo  {journal} {Phys. Rev. Lett.}\ }\textbf {\bibinfo
  {volume} {125}},\ \bibinfo {pages} {200501} (\bibinfo {year}
  {2020})}\BibitemShut {NoStop}%
\bibitem [{\citenamefont {Neven}\ \emph {et~al.}(2021)\citenamefont {Neven},
  \citenamefont {Carrasco}, \citenamefont {Vitale}, \citenamefont {Kokail},
  \citenamefont {Elben}, \citenamefont {Dalmonte}, \citenamefont {Calabrese},
  \citenamefont {Zoller}, \citenamefont {Vermersch}, \citenamefont {Kueng},\
  and\ \citenamefont {Kraus}}]{neven_symmetry-resolved_2021}%
  \BibitemOpen
  \bibfield  {author} {\bibinfo {author} {\bibfnamefont {A.}~\bibnamefont
  {Neven}}, \bibinfo {author} {\bibfnamefont {J.}~\bibnamefont {Carrasco}},
  \bibinfo {author} {\bibfnamefont {V.}~\bibnamefont {Vitale}}, \bibinfo
  {author} {\bibfnamefont {C.}~\bibnamefont {Kokail}}, \bibinfo {author}
  {\bibfnamefont {A.}~\bibnamefont {Elben}}, \bibinfo {author} {\bibfnamefont
  {M.}~\bibnamefont {Dalmonte}}, \bibinfo {author} {\bibfnamefont
  {P.}~\bibnamefont {Calabrese}}, \bibinfo {author} {\bibfnamefont
  {P.}~\bibnamefont {Zoller}}, \bibinfo {author} {\bibfnamefont
  {B.}~\bibnamefont {Vermersch}}, \bibinfo {author} {\bibfnamefont
  {R.}~\bibnamefont {Kueng}}, \ and\ \bibinfo {author} {\bibfnamefont
  {B.}~\bibnamefont {Kraus}},\ }\bibfield  {title} {\enquote {\bibinfo {title}
  {Symmetry-resolved entanglement detection using partial transpose moments},}\
  }\href {https://www.nature.com/articles/s41534-021-00487-y} {\bibfield
  {journal} {\bibinfo  {journal} {npj Quantum Inf.}\ }\textbf {\bibinfo
  {volume} {7}},\ \bibinfo {pages} {152} (\bibinfo {year} {2021})}\BibitemShut
  {NoStop}%
\bibitem [{\citenamefont {Rath}\ \emph {et~al.}(2021)\citenamefont {Rath},
  \citenamefont {Branciard}, \citenamefont {Minguzzi},\ and\ \citenamefont
  {Vermersch}}]{rath_quantum_2021}%
  \BibitemOpen
  \bibfield  {author} {\bibinfo {author} {\bibfnamefont {A.}~\bibnamefont
  {Rath}}, \bibinfo {author} {\bibfnamefont {C.}~\bibnamefont {Branciard}},
  \bibinfo {author} {\bibfnamefont {A.}~\bibnamefont {Minguzzi}}, \ and\
  \bibinfo {author} {\bibfnamefont {B.}~\bibnamefont {Vermersch}},\ }\bibfield
  {title} {\enquote {\bibinfo {title} {Quantum fisher information from
  randomized measurements},}\ }\href
  {https://link.aps.org/doi/10.1103/PhysRevLett.127.260501} {\bibfield
  {journal} {\bibinfo  {journal} {Phys. Rev. Lett.}\ }\textbf {\bibinfo
  {volume} {127}},\ \bibinfo {pages} {260501} (\bibinfo {year}
  {2021})}\BibitemShut {NoStop}%
\bibitem [{\citenamefont {Garcia}\ \emph {et~al.}(2021)\citenamefont {Garcia},
  \citenamefont {Zhou},\ and\ \citenamefont {Jaffe}}]{garcia_quantum_2021}%
  \BibitemOpen
  \bibfield  {author} {\bibinfo {author} {\bibfnamefont {R.~J.}\ \bibnamefont
  {Garcia}}, \bibinfo {author} {\bibfnamefont {Y.}~\bibnamefont {Zhou}}, \ and\
  \bibinfo {author} {\bibfnamefont {A.}~\bibnamefont {Jaffe}},\ }\bibfield
  {title} {\enquote {\bibinfo {title} {Quantum scrambling with classical
  shadows},}\ }\href
  {https://link.aps.org/doi/10.1103/PhysRevResearch.3.033155} {\bibfield
  {journal} {\bibinfo  {journal} {Phys. Rev. Research}\ }\textbf {\bibinfo
  {volume} {3}},\ \bibinfo {pages} {033155} (\bibinfo {year}
  {2021})}\BibitemShut {NoStop}%
\bibitem [{\citenamefont {Joshi}\ \emph {et~al.}(2022)\citenamefont {Joshi},
  \citenamefont {Elben}, \citenamefont {Vikram}, \citenamefont {Vermersch},
  \citenamefont {Galitski},\ and\ \citenamefont {Zoller}}]{joshi_probing_2022}%
  \BibitemOpen
  \bibfield  {author} {\bibinfo {author} {\bibfnamefont {L.~K.}\ \bibnamefont
  {Joshi}}, \bibinfo {author} {\bibfnamefont {A.}~\bibnamefont {Elben}},
  \bibinfo {author} {\bibfnamefont {A.}~\bibnamefont {Vikram}}, \bibinfo
  {author} {\bibfnamefont {B.}~\bibnamefont {Vermersch}}, \bibinfo {author}
  {\bibfnamefont {V.}~\bibnamefont {Galitski}}, \ and\ \bibinfo {author}
  {\bibfnamefont {P.}~\bibnamefont {Zoller}},\ }\bibfield  {title} {\enquote
  {\bibinfo {title} {Probing many-body quantum chaos with quantum
  simulators},}\ }\href {https://link.aps.org/doi/10.1103/PhysRevX.12.011018}
  {\bibfield  {journal} {\bibinfo  {journal} {Phys. Rev. X}\ }\textbf {\bibinfo
  {volume} {12}},\ \bibinfo {pages} {011018} (\bibinfo {year}
  {2022})}\BibitemShut {NoStop}%
\bibitem [{\citenamefont {Huang}\ \emph {et~al.}(2021)\citenamefont {Huang},
  \citenamefont {Kueng},\ and\ \citenamefont
  {Preskill}}]{huang_efficient_2021}%
  \BibitemOpen
  \bibfield  {author} {\bibinfo {author} {\bibfnamefont {H.~Y.}\ \bibnamefont
  {Huang}}, \bibinfo {author} {\bibfnamefont {R.}~\bibnamefont {Kueng}}, \ and\
  \bibinfo {author} {\bibfnamefont {J.}~\bibnamefont {Preskill}},\ }\bibfield
  {title} {\enquote {\bibinfo {title} {Efficient estimation of pauli
  observables by derandomization},}\ }\href
  {https://link.aps.org/doi/10.1103/PhysRevLett.127.030503} {\bibfield
  {journal} {\bibinfo  {journal} {Phys. Rev. Lett.}\ }\textbf {\bibinfo
  {volume} {127}},\ \bibinfo {pages} {030503} (\bibinfo {year}
  {2021})}\BibitemShut {NoStop}%
\bibitem [{\citenamefont {Zhang}\ \emph {et~al.}(2021)\citenamefont {Zhang},
  \citenamefont {Sun}, \citenamefont {Fang}, \citenamefont {Zhang},
  \citenamefont {Yuan},\ and\ \citenamefont
  {Lu}}]{zhang_experimental_measurement_shadows_2021}%
  \BibitemOpen
  \bibfield  {author} {\bibinfo {author} {\bibfnamefont {T.}~\bibnamefont
  {Zhang}}, \bibinfo {author} {\bibfnamefont {J.}~\bibnamefont {Sun}}, \bibinfo
  {author} {\bibfnamefont {X.~X.}\ \bibnamefont {Fang}}, \bibinfo {author}
  {\bibfnamefont {X.~M.}\ \bibnamefont {Zhang}}, \bibinfo {author}
  {\bibfnamefont {X.}~\bibnamefont {Yuan}}, \ and\ \bibinfo {author}
  {\bibfnamefont {H.}~\bibnamefont {Lu}},\ }\bibfield  {title} {\enquote
  {\bibinfo {title} {Experimental quantum state measurement with classical
  shadows},}\ }\href
  {https://journals.aps.org/prl/abstract/10.1103/PhysRevLett.127.200501}
  {\bibfield  {journal} {\bibinfo  {journal} {Phys. Rev. Lett.}\ }\textbf
  {\bibinfo {volume} {127}},\ \bibinfo {pages} {200501} (\bibinfo {year}
  {2021})}\BibitemShut {NoStop}%
\bibitem [{\citenamefont {Chen}\ \emph {et~al.}(2021)\citenamefont {Chen},
  \citenamefont {Yu}, \citenamefont {Zeng},\ and\ \citenamefont
  {Flammia}}]{chen_robust_shadow_estimation_2021}%
  \BibitemOpen
  \bibfield  {author} {\bibinfo {author} {\bibfnamefont {S.}~\bibnamefont
  {Chen}}, \bibinfo {author} {\bibfnamefont {W.}~\bibnamefont {Yu}}, \bibinfo
  {author} {\bibfnamefont {P.}~\bibnamefont {Zeng}}, \ and\ \bibinfo {author}
  {\bibfnamefont {S.~T.}\ \bibnamefont {Flammia}},\ }\bibfield  {title}
  {\enquote {\bibinfo {title} {Robust shadow estimation},}\ }\href
  {https://journals.aps.org/prxquantum/abstract/10.1103/PRXQuantum.2.030348}
  {\bibfield  {journal} {\bibinfo  {journal} {PRX Quantum}\ }\textbf {\bibinfo
  {volume} {2}},\ \bibinfo {pages} {030348} (\bibinfo {year}
  {2021})}\BibitemShut {NoStop}%
\bibitem [{\citenamefont {Hu}\ and\ \citenamefont
  {You}(2022)}]{hu_hamiltonian_driven_2022}%
  \BibitemOpen
  \bibfield  {author} {\bibinfo {author} {\bibfnamefont {H.~Y.}\ \bibnamefont
  {Hu}}\ and\ \bibinfo {author} {\bibfnamefont {Y.~Z.}\ \bibnamefont {You}},\
  }\bibfield  {title} {\enquote {\bibinfo {title} {Hamiltonian-driven shadow
  tomography of quantum states},}\ }\href
  {https://journals.aps.org/prresearch/abstract/10.1103/PhysRevResearch.4.013054}
  {\bibfield  {journal} {\bibinfo  {journal} {Phys. Rev. Research}\ }\textbf
  {\bibinfo {volume} {4}},\ \bibinfo {pages} {013054} (\bibinfo {year}
  {2022})}\BibitemShut {NoStop}%
\bibitem [{\citenamefont {Hu}\ \emph {et~al.}(2022)\citenamefont {Hu},
  \citenamefont {Choi},\ and\ \citenamefont
  {You}}]{hu_scrambled_dynamics_2022}%
  \BibitemOpen
  \bibfield  {author} {\bibinfo {author} {\bibfnamefont {H.~Y.}\ \bibnamefont
  {Hu}}, \bibinfo {author} {\bibfnamefont {S.}~\bibnamefont {Choi}}, \ and\
  \bibinfo {author} {\bibfnamefont {Y.~Z.}\ \bibnamefont {You}},\ }\bibfield
  {title} {\enquote {\bibinfo {title} {Classical shadow tomography with locally
  scrambled quantum dynamics},}\ }\href {https://arxiv.org/abs/2107.04817}
  {\bibfield  {journal} {\bibinfo  {journal} {arXiv:2107.04817}\ } (\bibinfo
  {year} {2022})}\BibitemShut {NoStop}%
\bibitem [{\citenamefont {Levy}\ \emph {et~al.}(2021)\citenamefont {Levy},
  \citenamefont {Luo},\ and\ \citenamefont {Clark}}]{levy_classical_2021}%
  \BibitemOpen
  \bibfield  {author} {\bibinfo {author} {\bibfnamefont {R.}~\bibnamefont
  {Levy}}, \bibinfo {author} {\bibfnamefont {D.}~\bibnamefont {Luo}}, \ and\
  \bibinfo {author} {\bibfnamefont {B.~K.}\ \bibnamefont {Clark}},\ }\bibfield
  {title} {\enquote {\bibinfo {title} {Classical shadows for quantum process
  tomography on near-term quantum computers},}\ }\href
  {http://arxiv.org/abs/2110.02965} {\bibfield  {journal} {\bibinfo  {journal}
  {arXiv:2110.02965}\ } (\bibinfo {year} {2021})}\BibitemShut {NoStop}%
\bibitem [{\citenamefont {Helsen}\ \emph {et~al.}(2021)\citenamefont {Helsen},
  \citenamefont {Ioannous}, \citenamefont {Roth}, \citenamefont {Kitzinger},
  \citenamefont {Onorati}, \citenamefont {Werner},\ and\ \citenamefont
  {Eisert}}]{helsen_estimating_2021}%
  \BibitemOpen
  \bibfield  {author} {\bibinfo {author} {\bibfnamefont {J.}~\bibnamefont
  {Helsen}}, \bibinfo {author} {\bibfnamefont {M.}~\bibnamefont {Ioannous}},
  \bibinfo {author} {\bibfnamefont {I.}~\bibnamefont {Roth}}, \bibinfo {author}
  {\bibfnamefont {J.}~\bibnamefont {Kitzinger}}, \bibinfo {author}
  {\bibfnamefont {E.}~\bibnamefont {Onorati}}, \bibinfo {author} {\bibfnamefont
  {A.~H.}\ \bibnamefont {Werner}}, \ and\ \bibinfo {author} {\bibfnamefont
  {J.}~\bibnamefont {Eisert}},\ }\bibfield  {title} {\enquote {\bibinfo {title}
  {Estimating gate-set properties from random sequences},}\ }\href
  {http://arxiv.org/abs/2110.13178} {\bibfield  {journal} {\bibinfo  {journal}
  {arXiv:2110.13178}\ } (\bibinfo {year} {2021})},\ \bibinfo {note} {arXiv:
  2110.13178}\BibitemShut {NoStop}%
\bibitem [{\citenamefont {Arute}\ \emph {et~al.}(2019)\citenamefont {Arute},
  \citenamefont {Arya}, \citenamefont {Babbush},\ and\ \citenamefont
  {et~al.}}]{arute_quantum_2019}%
  \BibitemOpen
  \bibfield  {author} {\bibinfo {author} {\bibfnamefont {F.}~\bibnamefont
  {Arute}}, \bibinfo {author} {\bibfnamefont {K.}~\bibnamefont {Arya}},
  \bibinfo {author} {\bibfnamefont {R.}~\bibnamefont {Babbush}}, \ and\
  \bibinfo {author} {\bibnamefont {et~al.}},\ }\bibfield  {title} {\enquote
  {\bibinfo {title} {Quantum supremacy using a programmable superconducting
  processor},}\ }\href {https://www.nature.com/articles/s41586-019-1666-5}
  {\bibfield  {journal} {\bibinfo  {journal} {Nature}\ }\textbf {\bibinfo
  {volume} {574}},\ \bibinfo {pages} {505–510} (\bibinfo {year}
  {2019})}\BibitemShut {NoStop}%
\bibitem [{\citenamefont {Chen}\ \emph {et~al.}(2019)\citenamefont {Chen},
  \citenamefont {Farahzad}, \citenamefont {Yoo},\ and\ \citenamefont
  {Wei}}]{Chen2019}%
  \BibitemOpen
  \bibfield  {author} {\bibinfo {author} {\bibfnamefont {Y.}~\bibnamefont
  {Chen}}, \bibinfo {author} {\bibfnamefont {M.}~\bibnamefont {Farahzad}},
  \bibinfo {author} {\bibfnamefont {S.}~\bibnamefont {Yoo}}, \ and\ \bibinfo
  {author} {\bibfnamefont {T.-C.}\ \bibnamefont {Wei}},\ }\bibfield  {title}
  {\enquote {\bibinfo {title} {Detector tomography on {IBM} quantum computers
  and mitigation of an imperfect measurement},}\ }\href
  {https://link.aps.org/doi/10.1103/PhysRevA.100.052315} {\bibfield  {journal}
  {\bibinfo  {journal} {Phys. Rev. A}\ }\textbf {\bibinfo {volume} {100}},\
  \bibinfo {pages} {052315} (\bibinfo {year} {2019})}\BibitemShut {NoStop}%
\bibitem [{\citenamefont {Acharya}\ \emph {et~al.}(2021)\citenamefont
  {Acharya}, \citenamefont {Saha},\ and\ \citenamefont
  {Sengupta}}]{acharya_journal_2022}%
  \BibitemOpen
  \bibfield  {author} {\bibinfo {author} {\bibfnamefont {A.}~\bibnamefont
  {Acharya}}, \bibinfo {author} {\bibfnamefont {S.}~\bibnamefont {Saha}}, \
  and\ \bibinfo {author} {\bibfnamefont {A.~M.}\ \bibnamefont {Sengupta}},\
  }\bibfield  {title} {\enquote {\bibinfo {title} {Shadow tomography based on
  informationally complete positive operator-valued measure},}\ }\href
  {\doibase 10.1103/PhysRevA.104.052418} {\bibfield  {journal} {\bibinfo
  {journal} {Phys. Rev. A}\ }\textbf {\bibinfo {volume} {104}},\ \bibinfo
  {pages} {052418} (\bibinfo {year} {2021})}\BibitemShut {NoStop}%
\bibitem [{sup()}]{supp}%
  \BibitemOpen
  \href@noop {} {\ }\bibinfo {note} {Supplementary Material, which contains the
  derivation of the classical shadow from the least square estimator, the
  discussion on the relationship between randomising unitaries with generalised
  measurements, the symmetry analysis of classical shadows, the proof of the
  optimality of the octahedron measurement for the construction of random
  observables with further
  references~\cite{teiko,DAriano_2005,unfolding_noise_2020,error_circuit_2017,Press1994,Granville1994,schurweylduality}}\BibitemShut
  {NoStop}%
\bibitem [{Note1()}]{Note1}%
  \BibitemOpen
  \bibinfo {note} {In practice, the targeted state could be very different from
  the worst case scenario assumed in obtaining the shadow norm. Therefore, it
  might also be informative to consider the average of the variance with
  respect to certain ensemble of states.}\BibitemShut {Stop}%
\bibitem [{\citenamefont {Bu}\ \emph {et~al.}(2022)\citenamefont {Bu},
  \citenamefont {Koh}, \citenamefont {Garcia},\ and\ \citenamefont
  {Jaffe}}]{bu_classical_2022}%
  \BibitemOpen
  \bibfield  {author} {\bibinfo {author} {\bibfnamefont {K.}~\bibnamefont
  {Bu}}, \bibinfo {author} {\bibfnamefont {D.~E.}\ \bibnamefont {Koh}},
  \bibinfo {author} {\bibfnamefont {R.~J.}\ \bibnamefont {Garcia}}, \ and\
  \bibinfo {author} {\bibfnamefont {A.}~\bibnamefont {Jaffe}},\ }\bibfield
  {title} {\enquote {\bibinfo {title} {Classical shadows with {Pauli}-invariant
  unitary ensembles},}\ }\href {http://arxiv.org/abs/2202.03272} {\bibfield
  {journal} {\bibinfo  {journal} {arXiv:2202.03272}\ } (\bibinfo {year}
  {2022})}\BibitemShut {NoStop}%
\bibitem [{\citenamefont {Nguyen}\ \emph {et~al.}(2020)\citenamefont {Nguyen},
  \citenamefont {Designolle}, \citenamefont {Barakat},\ and\ \citenamefont
  {Gühne}}]{nguyen_symmetries_2020}%
  \BibitemOpen
  \bibfield  {author} {\bibinfo {author} {\bibfnamefont {H.~C.}\ \bibnamefont
  {Nguyen}}, \bibinfo {author} {\bibfnamefont {S.}~\bibnamefont {Designolle}},
  \bibinfo {author} {\bibfnamefont {M.}~\bibnamefont {Barakat}}, \ and\
  \bibinfo {author} {\bibfnamefont {O.}~\bibnamefont {Gühne}},\ }\bibfield
  {title} {\enquote {\bibinfo {title} {Symmetries between measurements in
  quantum mechanics},}\ }\href {https://arxiv.org/abs/2003.12553} {\bibfield
  {journal} {\bibinfo  {journal} {arXiv:2003.12553}\ } (\bibinfo {year}
  {2020})}\BibitemShut {NoStop}%
\bibitem [{\citenamefont {Zhu}\ and\ \citenamefont
  {B.~G.~Englert}(2011)}]{zhu_tomography_2011}%
  \BibitemOpen
  \bibfield  {author} {\bibinfo {author} {\bibfnamefont {H.}~\bibnamefont
  {Zhu}}\ and\ \bibinfo {author} {\bibfnamefont {B.-G.}\ \bibnamefont
  {B.~G.~Englert}},\ }\bibfield  {title} {\enquote {\bibinfo {title} {Quantum
  state tomography with fully symmetric measurements and product
  measurements},}\ }\href {\doibase 10.1103/PhysRevA.84.022327} {\bibfield
  {journal} {\bibinfo  {journal} {Phys. Rev. A}\ }\textbf {\bibinfo {volume}
  {84}},\ \bibinfo {pages} {022327} (\bibinfo {year} {2011})}\BibitemShut
  {NoStop}%
\bibitem [{\citenamefont {Bogdanov}\ \emph {et~al.}(2011)\citenamefont
  {Bogdanov}, \citenamefont {Brida}, \citenamefont {Bukeev}, \citenamefont
  {Genovese}, \citenamefont {Kravtsov}, \citenamefont {Kulik}, \citenamefont
  {Moreva}, \citenamefont {Soloviev},\ and\ \citenamefont
  {Shurupov}}]{bogdanov_statistical_2011}%
  \BibitemOpen
  \bibfield  {author} {\bibinfo {author} {\bibfnamefont {Y.~I.}\ \bibnamefont
  {Bogdanov}}, \bibinfo {author} {\bibfnamefont {G.}~\bibnamefont {Brida}},
  \bibinfo {author} {\bibfnamefont {I.~D.}\ \bibnamefont {Bukeev}}, \bibinfo
  {author} {\bibfnamefont {M.}~\bibnamefont {Genovese}}, \bibinfo {author}
  {\bibfnamefont {K.~S.}\ \bibnamefont {Kravtsov}}, \bibinfo {author}
  {\bibfnamefont {S.~P.}\ \bibnamefont {Kulik}}, \bibinfo {author}
  {\bibfnamefont {E.~V.}\ \bibnamefont {Moreva}}, \bibinfo {author}
  {\bibfnamefont {A.~A.}\ \bibnamefont {Soloviev}}, \ and\ \bibinfo {author}
  {\bibfnamefont {A.~P.}\ \bibnamefont {Shurupov}},\ }\bibfield  {title}
  {\enquote {\bibinfo {title} {Statistical estimation of the quality of
  quantum-tomography protocols},}\ }\href {\doibase 10.1103/PhysRevA.84.042108}
  {\bibfield  {journal} {\bibinfo  {journal} {Phys. Rev. A}\ }\textbf {\bibinfo
  {volume} {84}},\ \bibinfo {pages} {042108} (\bibinfo {year}
  {2011})}\BibitemShut {NoStop}%
\bibitem [{\citenamefont {Bravyi}\ \emph {et~al.}(2021)\citenamefont {Bravyi},
  \citenamefont {Sheldon}, \citenamefont {Kandala}, \citenamefont {Mckay},\
  and\ \citenamefont {Gambetta}}]{error_multiqubits_2021}%
  \BibitemOpen
  \bibfield  {author} {\bibinfo {author} {\bibfnamefont {S.}~\bibnamefont
  {Bravyi}}, \bibinfo {author} {\bibfnamefont {S.}~\bibnamefont {Sheldon}},
  \bibinfo {author} {\bibfnamefont {A.}~\bibnamefont {Kandala}}, \bibinfo
  {author} {\bibfnamefont {D.~C.}\ \bibnamefont {Mckay}}, \ and\ \bibinfo
  {author} {\bibfnamefont {J.~M.}\ \bibnamefont {Gambetta}},\ }\bibfield
  {title} {\enquote {\bibinfo {title} {Mitigating measurement errors in
  multiqubit experiments},}\ }\href {\doibase 10.1103/PhysRevA.103.042605}
  {\bibfield  {journal} {\bibinfo  {journal} {Phys. Rev. A}\ }\textbf {\bibinfo
  {volume} {103}},\ \bibinfo {pages} {042605} (\bibinfo {year}
  {2021})}\BibitemShut {NoStop}%
\bibitem [{\citenamefont {Hicks}\ \emph {et~al.}(2022)\citenamefont {Hicks},
  \citenamefont {Kobrin}, \citenamefont {Bauer},\ and\ \citenamefont
  {Nachman}}]{active_error_2022}%
  \BibitemOpen
  \bibfield  {author} {\bibinfo {author} {\bibfnamefont {R.}~\bibnamefont
  {Hicks}}, \bibinfo {author} {\bibfnamefont {B.}~\bibnamefont {Kobrin}},
  \bibinfo {author} {\bibfnamefont {C.~W.}\ \bibnamefont {Bauer}}, \ and\
  \bibinfo {author} {\bibfnamefont {B.}~\bibnamefont {Nachman}},\ }\bibfield
  {title} {\enquote {\bibinfo {title} {Active readout-error mitigation},}\
  }\href {\doibase 10.1103/PhysRevA.105.012419} {\bibfield  {journal} {\bibinfo
   {journal} {Phys. Rev. A}\ }\textbf {\bibinfo {volume} {105}},\ \bibinfo
  {pages} {012419} (\bibinfo {year} {2022})}\BibitemShut {NoStop}%
\bibitem [{\citenamefont {Fischer}\ \emph {et~al.}(2022)\citenamefont
  {Fischer}, \citenamefont {Miller}, \citenamefont {Tacchino}, \citenamefont
  {Barkoutsos}, \citenamefont {Egger},\ and\ \citenamefont
  {Tavernelli}}]{fischer_ancilla_free_2022}%
  \BibitemOpen
  \bibfield  {author} {\bibinfo {author} {\bibfnamefont {L.~E.}\ \bibnamefont
  {Fischer}}, \bibinfo {author} {\bibfnamefont {D.}~\bibnamefont {Miller}},
  \bibinfo {author} {\bibfnamefont {F.}~\bibnamefont {Tacchino}}, \bibinfo
  {author} {\bibfnamefont {P.~Kl.}\ \bibnamefont {Barkoutsos}}, \bibinfo
  {author} {\bibfnamefont {D.~J.}\ \bibnamefont {Egger}}, \ and\ \bibinfo
  {author} {\bibfnamefont {I.}~\bibnamefont {Tavernelli}},\ }\bibfield  {title}
  {\enquote {\bibinfo {title} {Ancilla-free implementation of generalized
  measurements for qubits embedded in a qudit space},}\ }\href
  {https://arxiv.org/abs/2203.07369} {\bibfield  {journal} {\bibinfo  {journal}
  {arXiv:2203.07369}\ } (\bibinfo {year} {2022})}\BibitemShut {NoStop}%
\bibitem [{\citenamefont {Stricker}\ \emph {et~al.}(2022)\citenamefont
  {Stricker}, \citenamefont {Meth}, \citenamefont {Postler}, \citenamefont
  {Edmunds}, \citenamefont {Ferrie}, \citenamefont {Blatt}, \citenamefont
  {Schindler}, \citenamefont {Monz}, \citenamefont {Kueng},\ and\ \citenamefont
  {Ringbauer}}]{stricker_2022}%
  \BibitemOpen
  \bibfield  {author} {\bibinfo {author} {\bibfnamefont {R.}~\bibnamefont
  {Stricker}}, \bibinfo {author} {\bibfnamefont {M.}~\bibnamefont {Meth}},
  \bibinfo {author} {\bibfnamefont {L.}~\bibnamefont {Postler}}, \bibinfo
  {author} {\bibfnamefont {C.}~\bibnamefont {Edmunds}}, \bibinfo {author}
  {\bibfnamefont {C.}~\bibnamefont {Ferrie}}, \bibinfo {author} {\bibfnamefont
  {R.}~\bibnamefont {Blatt}}, \bibinfo {author} {\bibfnamefont
  {P.}~\bibnamefont {Schindler}}, \bibinfo {author} {\bibfnamefont
  {T.}~\bibnamefont {Monz}}, \bibinfo {author} {\bibfnamefont {R.}~\bibnamefont
  {Kueng}}, \ and\ \bibinfo {author} {\bibfnamefont {M.}~\bibnamefont
  {Ringbauer}},\ }\bibfield  {title} {\enquote {\bibinfo {title} {Experimental
  single-setting quantum state tomography},}\ }\href
  {https://arxiv.org/abs/2206.00019} {\bibfield  {journal} {\bibinfo  {journal}
  {arXiv:2206.00019}\ } (\bibinfo {year} {2022})}\BibitemShut {NoStop}%
\bibitem [{\citenamefont {McNulty}\ \emph {et~al.}(2022)\citenamefont
  {McNulty}, \citenamefont {Maciejewski},\ and\ \citenamefont
  {Oszmaniec}}]{mcnulty_2022}%
  \BibitemOpen
  \bibfield  {author} {\bibinfo {author} {\bibfnamefont {D.}~\bibnamefont
  {McNulty}}, \bibinfo {author} {\bibfnamefont {F.~B.}\ \bibnamefont
  {Maciejewski}}, \ and\ \bibinfo {author} {\bibfnamefont {M.}~\bibnamefont
  {Oszmaniec}},\ }\bibfield  {title} {\enquote {\bibinfo {title} {Estimating
  quantum hamiltonians via joint measurements of noisy non-commuting
  observables},}\ }\href {https://arxiv.org/abs/2206.08912} {\bibfield
  {journal} {\bibinfo  {journal} {arXiv:2206.08912}\ } (\bibinfo {year}
  {2022})}\BibitemShut {NoStop}%
\bibitem [{\citenamefont {Heinosaari}\ and\ \citenamefont
  {Ziman}(2012)}]{teiko}%
  \BibitemOpen
  \bibfield  {author} {\bibinfo {author} {\bibfnamefont {T.}~\bibnamefont
  {Heinosaari}}\ and\ \bibinfo {author} {\bibfnamefont {M.}~\bibnamefont
  {Ziman}},\ }\href@noop {} {\emph {\bibinfo {title} {The Mathematical Language
  of Quantum Theory: From Uncertainty to Entanglement}}}\ (\bibinfo
  {publisher} {Cambridge University Press},\ \bibinfo {year}
  {2012})\BibitemShut {NoStop}%
\bibitem [{\citenamefont {D{\textquotesingle}Ariano}\ \emph
  {et~al.}(2005)\citenamefont {D{\textquotesingle}Ariano}, \citenamefont
  {Presti},\ and\ \citenamefont {Perinotti}}]{DAriano_2005}%
  \BibitemOpen
  \bibfield  {author} {\bibinfo {author} {\bibfnamefont {G.~M.}\ \bibnamefont
  {D{\textquotesingle}Ariano}}, \bibinfo {author} {\bibfnamefont {P.~Lo}\
  \bibnamefont {Presti}}, \ and\ \bibinfo {author} {\bibfnamefont
  {P.}~\bibnamefont {Perinotti}},\ }\bibfield  {title} {\enquote {\bibinfo
  {title} {Classical randomness in quantum measurements},}\ }\href {\doibase
  10.1088/0305-4470/38/26/010} {\bibfield  {journal} {\bibinfo  {journal} {J.
  Phys. A: Math. Gen.}\ }\textbf {\bibinfo {volume} {38}},\ \bibinfo {pages}
  {5979--5991} (\bibinfo {year} {2005})}\BibitemShut {NoStop}%
\bibitem [{\citenamefont {Nachman}\ \emph {et~al.}(2020)\citenamefont
  {Nachman}, \citenamefont {Urbanek}, \citenamefont {de~Jong},\ and\
  \citenamefont {Bauer}}]{unfolding_noise_2020}%
  \BibitemOpen
  \bibfield  {author} {\bibinfo {author} {\bibfnamefont {B.}~\bibnamefont
  {Nachman}}, \bibinfo {author} {\bibfnamefont {M.}~\bibnamefont {Urbanek}},
  \bibinfo {author} {\bibfnamefont {W.~A.}\ \bibnamefont {de~Jong}}, \ and\
  \bibinfo {author} {\bibfnamefont {C.~W.}\ \bibnamefont {Bauer}},\ }\bibfield
  {title} {\enquote {\bibinfo {title} {Unfolding quantum computer readout
  noise},}\ }\href {https://www.nature.com/articles/s41534-020-00309-7}
  {\bibfield  {journal} {\bibinfo  {journal} {npj Quantum Inf.}\ }\textbf
  {\bibinfo {volume} {6}},\ \bibinfo {pages} {84} (\bibinfo {year}
  {2020})}\BibitemShut {NoStop}%
\bibitem [{\citenamefont {Temme}\ \emph {et~al.}(2017)\citenamefont {Temme},
  \citenamefont {Bravyi},\ and\ \citenamefont {Gambetta}}]{error_circuit_2017}%
  \BibitemOpen
  \bibfield  {author} {\bibinfo {author} {\bibfnamefont {K.}~\bibnamefont
  {Temme}}, \bibinfo {author} {\bibfnamefont {S.}~\bibnamefont {Bravyi}}, \
  and\ \bibinfo {author} {\bibfnamefont {J.~M.}\ \bibnamefont {Gambetta}},\
  }\bibfield  {title} {\enquote {\bibinfo {title} {Error mitigation for
  short-depth quantum circuits},}\ }\href {\doibase
  10.1103/PhysRevLett.119.180509} {\bibfield  {journal} {\bibinfo  {journal}
  {Phys. Rev. Lett.}\ }\textbf {\bibinfo {volume} {119}},\ \bibinfo {pages}
  {180509} (\bibinfo {year} {2017})}\BibitemShut {NoStop}%
\bibitem [{\citenamefont {Press}\ \emph {et~al.}(2007)\citenamefont {Press},
  \citenamefont {Teukolsky}, \citenamefont {Vetterling},\ and\ \citenamefont
  {Flannery}}]{Press1994}%
  \BibitemOpen
  \bibfield  {author} {\bibinfo {author} {\bibfnamefont {W.~H.}\ \bibnamefont
  {Press}}, \bibinfo {author} {\bibfnamefont {S.~A.}\ \bibnamefont
  {Teukolsky}}, \bibinfo {author} {\bibfnamefont {W.~T.}\ \bibnamefont
  {Vetterling}}, \ and\ \bibinfo {author} {\bibfnamefont {B.~P.}\ \bibnamefont
  {Flannery}},\ }\href@noop {} {\emph {\bibinfo {title} {Numerical recipes}}}\
  (\bibinfo  {publisher} {Cambridge University Press},\ \bibinfo {year}
  {2007})\BibitemShut {NoStop}%
\bibitem [{\citenamefont {Granville}\ \emph {et~al.}(1994)\citenamefont
  {Granville}, \citenamefont {Krivanek},\ and\ \citenamefont
  {Rasson}}]{Granville1994}%
  \BibitemOpen
  \bibfield  {author} {\bibinfo {author} {\bibfnamefont {V.}~\bibnamefont
  {Granville}}, \bibinfo {author} {\bibfnamefont {M.}~\bibnamefont {Krivanek}},
  \ and\ \bibinfo {author} {\bibfnamefont {J.-P.}\ \bibnamefont {Rasson}},\
  }\bibfield  {title} {\enquote {\bibinfo {title} {Simulated annealing: a proof
  of convergence},}\ }\href {\doibase 10.1109/34.295910} {\bibfield  {journal}
  {\bibinfo  {journal} {IEEE Trans. Pattern Anal. Mach. Intell.}\ }\textbf
  {\bibinfo {volume} {16}},\ \bibinfo {pages} {652--656} (\bibinfo {year}
  {1994})}\BibitemShut {NoStop}%
\bibitem [{\citenamefont {Fulton}\ and\ \citenamefont
  {Harris}(2004)}]{schurweylduality}%
  \BibitemOpen
  \bibfield  {author} {\bibinfo {author} {\bibfnamefont {W.}~\bibnamefont
  {Fulton}}\ and\ \bibinfo {author} {\bibfnamefont {J.}~\bibnamefont
  {Harris}},\ }\href {https://link.springer.com/book/10.1007/978-1-4612-0979-9}
  {\emph {\bibinfo {title} {Representation Theory}}}\ (\bibinfo  {publisher}
  {Springer New York},\ \bibinfo {year} {2004})\BibitemShut {NoStop}%
\end{thebibliography}%

\newpage
\newpage



\section*{Supplementary material}

# Supplementary Material:
## Optimising shadow tomography with generalised measurements

H. Chau Nguyen, Jan Lennart Bönsel, Jonathan Steinberg, and Otfried Gühne
*Naturwissenschaftlich–Technische Fakultät, Universität Siegen,
Walter-Flex-Straße 3, 57068 Siegen, Germany*
(Dated: November 24, 2022)

### Appendix A: Least square derivation of shadow tomography with generalised measurements

The derivation of the solution to the least square problem is standard in data science and machine learning literature [1], which has also been used in the theory of quantum tomography, see e.g., Ref. [2]. For a self-contained complete understanding of the logic, we present here the derivation specifically for shadow tomography.

Recall that a measurement $E = \{E_k\}_{k=1}^{N}$ is associated with a map $\Phi_E$ which maps $\rho$ to a probability distribution, $\Phi_E : M_D \to \mathbb{R}^N$, $\Phi_E(\rho) = \{\mathrm{Tr}(\rho E_k)\}_{k=1}^{N}$. An unbiased linear estimator for $\rho$ is a map $\chi : \mathbb{R}^N \to M_D$ which maps a probability distribution to a quantum state such that $(\chi \circ \Phi_E)$ acts as identity over $M_D$. That means, as long as the statistics $\Phi_E(\rho)$ of a state $\rho$ can be exactly measured, $\chi$ allows for an exact construction of the state $\rho$. Note that if $\{E_k\}_{k=1}^{N}$ form a basis for the operator space, that is, $E$ is informationally complete, but not overcomplete, then $\Phi_E$ is invertible and $\chi = \Phi_E^{-1}$. In the case that $E$ is informationally *overcomplete*, $\chi$ is not uniquely defined. In fact, $\Phi_E$ is not surjective in this case: there are probability distributions in $\mathbb{R}^N$ that give rise to no state $\rho$. In this situation, the natural choice is the least square estimator, namely, for a distribution $\vec{p} \in \mathbb{R}^N$, the estimated state is defined by

$$\chi_{\mathrm{LS}}(\vec{p}) = \arg\min_{\tau} L(\tau), \tag{A1}$$

where

$$L(\tau) = \sum_{i=1}^{N} [\mathrm{Tr}(\tau E_i) - p_i]^2. \tag{A2}$$

To carry out this minimisation, we consider the expansion of $\delta L = L(\tau + \delta\tau) - L(\tau)$ in the first order of the small variation $\delta\tau$,

$$\delta L = 2 \sum_{i=1}^{N} [\mathrm{Tr}(\tau E_i) - p_i]\, \mathrm{Tr}(E_i \delta\tau). \tag{A3}$$

That $L$ is minimised implies that $\delta L = 0$ for all choices of $\delta\tau$. It must then hold that

$$\sum_{i=1}^{N} [\mathrm{Tr}(\tau E_i) - p_i] E_i = 0. \tag{A4}$$

Using the definition of $\Phi_E$, one can rewrite the above equation as

$$\Phi_E^{\dagger}[\Phi_E(\tau) - \vec{p}] = 0. \tag{A5}$$

The solution of the estimator $\chi_{\mathrm{LS}}(\vec{p})$ can be found to be

$$\chi_{\mathrm{LS}} = (\Phi_E^{\dagger}\Phi_E)^{-1}\Phi_E^{\dagger}. \tag{A6}$$

Observe that when $\Phi_E$ is invertible, $\chi_{\mathrm{LS}} = \Phi_E^{-1}$.

This procedure gives the exact reconstruction of $\rho$ as long as the exact (thus valid) statistics $\vec{p}$ for the measurement outcomes is available. Remarkably this estimator is linear, which implies that the estimated state over the whole data set can be split into the sum of the estimated states for the single data points. Indeed, a single observation of an outcome $k$ can be associated with an elementary statistics vector $\vec{q}_k = \{\delta_{ki}\}_{i=1}^{N} \in \mathbb{R}^N$. The statistics of the whole dataset of a string of outcomes $\{k_i\}_{i=1}^{M}$ is given by

$$\vec{p} = \frac{1}{M} \sum_{i=1}^{M} \vec{q}_{k_i}, \tag{A7}$$

and therefore

$$\chi_{\mathrm{LS}}(\vec{p}) = \frac{1}{M} \sum_{i=1}^{M} \chi_{\mathrm{LS}}(\vec{q}_{k_i}). \tag{A8}$$

One can then confirm that $\chi_{\mathrm{LS}}(\vec{q}_{k_i})$ is exactly the classical shadow corresponding to the outcome $k_i$ as given in Eq. (3) in the main text.

Notice that, in similarity to Ref. [3], it is also easy to see that over an infinite number of repeats of the measurement, the average of the classical shadows converges to $\rho$, i.e., if $k_i \in \{1, ..., N\}$ denotes the measurement outcome of the $i$-th repeat, we have $\lim_{M \to \infty} \frac{1}{M} \sum_{i=1}^{M} \hat{\rho}_{k_i} = \rho$. It is important to emphasise, however, that this convergence does *not* guarantee that $C_E^{-1}(E_k)$ is a good estimator [4]. That $C_E^{-1}(E_k)$ is a good estimator is supported by the fact that it is actually the least square estimator. It should also be noted that, despite being an unbiased estimator, in general, $\hat{\rho}_k$ is not unit trace; this is only the case if all the effects $E_k$ share the same trace.

The linearity of the estimator is crucial to shadow tomography. For instance, this allows one to estimate linear observables with single data points and later on average over the whole dataset. Although in the low sampling regime, $M \ll D$, the estimated state can be far from the

targeted state in the state space, it can be sufficient to estimate many linear observables [3, 5]. The name (classical) shadow tomography reflects the fact that we are not aiming at reconstructing the density operator of the system, but only the classical values of observables [5]. For more on the rationale and detailed theory behind, we refer the readers to Ref. [5] and the references therein.

## Appendix B: Protocol of shadow tomography with generalised measurements

Let us summary the step-by-step protocol of shadow tomography with generalised measurement $E = \{E_i\}_{i=1}^N$ for the estimation of a set of observables $\mathcal{X} = \{X_i\}_{i=1}^m$.

1. Given the measurement $E = \{E_i\}_{i=1}^N$, classical shadows $\{\hat{\rho}_k\}_{k=1}^N$ are classically computed using formula eq. (3) or eq. (6) in the main text.

2. The quantum system is prepared in the designed state for investigation and the measurement $E$ is carried out. This is repeated on the system $M$ times and the string of outcomes $\{k_j\}_{j=1}^M$ is recorded.

3a. The mean values of $\{X_i\}_{i=1}^m$ are estimated by $\langle X_i \rangle \approx 1/M \sum_{j=1}^M \text{Tr}(\hat{\rho}_{k_j} X_i)$.

Although not discussed in the main text, the above procedure can be modified to estimate polynomial functions of $\rho$ as described in Ref. [3]. For example, to estimate the purity $\text{Tr}(\rho^2)$, the last step is simply replaced by

3b. The linear entropy $\text{Tr}(\rho^2)$ is estimated by $\text{Tr}(\rho^2) \approx 1/[M(M-1)] \sum_{j_1 \neq j_2} \text{Tr}(\hat{\rho}_{j_1} \hat{\rho}_{j_2})$.

## Appendix C: Independence of post-measurement states

Our suggested classical shadows are different from the suggestion from Ref. [6]. In order to apply the original formulation of classical shadow of Ref. [3], Ref. [6] suggested to synthesise the post-measurement states of the system after the generalised measurement. Note that post-measurement states are not available to a generalised measurement *per se*, but depends on the instrument that realises the measurement. It was suggested [6] that one manually chooses the eigenstates of the measurement effect $E_k$ of the obtained outcome $k$ according to a chosen probability distribution as the post-measurement states. In practice, it was suggested to choose the eigenvectors of highest eigenvalues of the effects as the post-measurement states. With the post-measurement states, construction of classical shadows is carried out using that of Ref. [3].

As we emphasised, our derivation of the classical shadows requires *no* post-measurement states of the generalised measurements. This frees the classical shadows from dependence on the choice of the probability distribution required to synthesise the post-measurement states as mentioned above. Moreover, it also gives a general derivation of classical shadows, containing that of Ref. [3] as a special case. As we shown in the main text, our simple generalisation of shadow tomography is crucial to the ability to optimise it in practice.

## Appendix D: Relation to randomised ideal measurements

The often used scheme of randomised ideal projective measurements as described in Ref. [3] can be considered as a special realisation of a generalised measurement. Specifically, following Ref. [3], a random unitary $U$ is drawn from a selected set $\mathcal{U}$ of unitaries and carried out on the system before making the measurement in a standard basis $\{|b\rangle\}_{b=1}^D$ (such as the computational basis). This corresponds to the random projective measurement in the basis $\{U^\dagger |b\rangle\}_{b=1}^D$. Then the whole procedure effectively simulates a generalised measurement with effects $\{1/|\mathcal{U}| U^\dagger |b\rangle\langle b| U : U \in \mathcal{U}, b = 1, 2, \ldots, D\}$, where $|\mathcal{U}|$ is the total number of unitaries in $\mathcal{U}$. Mathematically, the resulted generalised measurement is a convex combination of the chosen ideal measurements. This is a well-known method of simulating generalised measurements with ideal ones, known as preprocessing [7].

Our suggested scheme of implementing shadow tomography using a single generalised measurement brings several interesting new perspectives. Firstly, the randomised unitary description is heavily over parametrised, i.e., having much more parameters than necessary. For example, the unitary ensemble of the Clifford group on a qubit contains 24 elements, and yet is equivalent to a single generalised measurement of 6 outcomes corresponding to six directions of the three standard axes, $1/3 \times \{|x^\pm\rangle\langle x^\pm|, |y^\pm\rangle\langle y^\pm|, |z^\pm\rangle\langle z^\pm|\}$. This over parametrisation makes it difficult to keep track over the parameters and to optimise the procedure. Furthermore, the same generalised measurement can be simulated with different sets of unitaries and there are also generalised measurements that cannot be simulated by unitary ensembles (such as the tetrahedron measurement discussed in the main text) [8]. With all these difficulties, finding an optimal set of unitaries for shadow tomography appears to be intractable, even for the simplest case of a single qubit. As we have showed in the main text, using generalised measurements, the problem of optimising shadow tomography turns out to be possible.

Using generalised measurements also brings a new perspective on the implementation of the procedure. A generalised measurement can be implemented by a measurement on the ancillary system after an appropriate coupling to the objective system. While this can still be

challenging in practice, it avoids changing measurement settings, which requires frequent (re)calibration of the setup.

### Appendix E: Symmetry of generalised measurements and the computation of classical shadows

In this appendix, we describe how the symmetry of a generalised measurement can be analysed and how the formula for the classical shadow can be drawn just from the consideration of symmetry. We say that a measurement $E = (E_1, E_2, \ldots, E_N)$ is symmetric if there is a subgroup $G$ of the permutation group over $(1, 2, \ldots, N)$ and a representation $U : G \to U(D)$ such that $E_{g(k)} = U_g E_k U_{g^{-1}}$ [9]. One can easily show that if $E$ is symmetric under $G$, then also $\Phi_E : M_D \to \mathbb{R}^N$ is covariant under $G$, that is, $\Phi_E(U_g \rho U_{g^{-1}}) = g[\Phi_E(\rho)]$, where $g$ acts on a probability distribution $\vec{p}$ by $[g(\vec{p})]_k = p_{g^{-1}(k)}$. As a consequence, $\Phi_E^\dagger$ is covariant under $G$, i.e., $\Phi_E^\dagger[g(\vec{p})] = U_g \Phi^\dagger(\vec{p}) U_{g^{-1}}$ for all distribution $\vec{p}$. This in turn implies that $C_E = \Phi_E^\dagger \Phi_E$ and particularly its inverse $C_E^{-1}$ are also covariant under $G$, $C_E^{-1}(U_g X U_{g^{-1}}) = U_g C_E^{-1}(X) U_{g^{-1}}$ for all hermitian operators $X$.

We consider the case that $E$ is uniform, which means $G$ acts transitively on the outcomes, or in other words any effect can be mapped to any other effect. In particular, this implies that $\text{Tr}(E_k)$ is independent of $k$, and we denote $\alpha = \text{Tr}(E_k)$. Following Ref. [9], we consider the stabiliser subgroup $G_k$ over $k$, that is, the subgroup of $G$ that leaves $k$ invariant. Since $G_k$ leaves $k$ invariant, $E_k$ commutes with $U(G_k)$. Because $C_E^{-1}$ is covariant under $G$, it is clear that $\hat{\rho}_k = C_E^{-1}(E_k)$ also commutes with $U(G_k)$. One says that $E$ is rigidly symmetric [9] if the representation $U$ restricted to any stabiliser subgroup over $k$, $G_k$, has exactly two irreducible representations. This is the characterisation of the property that, an operator commuting with $U(G_k)$ can only be a linear combination of $E_k$ and $\mathbb{1}$. Hence if $E$ is rigidly symmetric, then

$$\hat{\rho}_k = a E_k + b \mathbb{1}. \tag{E1}$$

To compute $a$ and $b$, one notice that $C_E(\hat{\rho}_k) = a \sum_{l=1}^N \text{Tr}(E_k E_l) E_l + b\alpha \mathbb{1}$, which is identified with $E_k$. Taking the trace of this equation and the trace of it after multiplying the two sides with $E_k$, one obtains

$$\alpha = a\alpha^2 + b\alpha D, \tag{E2}$$
$$\beta = a\gamma + b\alpha^2, \tag{E3}$$

with $\beta = \text{Tr}(E_k^2)$, $\gamma = \sum_{l=1}^N \text{Tr}(E_k E_l)^2$, both independent of $k$ due to the uniformity. The coefficients $a$ and $b$ can be then explicitly solved,

$$a = (D\beta - \alpha^2)/(D\gamma - \alpha^3), \tag{E4}$$
$$b = (\gamma - \alpha\beta)/(D\gamma - \alpha^3). \tag{E5}$$

| $d$ | ST | $N$ | Comments | $a$ | $b$ |
|---|---|---|---|---|---|
| 2 | 8 | 6 | Octahedron | 9 | $-1$ |
| | | 8 | Cube | 12 | $-1$ |
| | | 12 | Cuboctahedron | 18 | $-1$ |
| | 16 | 12 | Icosahedron | 18 | $-1$ |
| | | 20 | Dodecahedron | 30 | $-1$ |
| | | 30 | Icosidodecahedron | 45 | $-1$ |
| 3 | 24 | 21 | | 28 | $-1$ |
| | 25 | 12 | csMUB | 16 | $-1$ |
| | 27 | 45 | | 60 | $-1$ |
| | | 60 | | 80 | $-1$ |
| 4 | 28 | 12 | Real MUB | 9 | $-\frac{1}{2}$ |
| | 29 | 20 | csMUB | 25 | $-1$ |
| | | 40 | | 50 | $-1$ |
| | | 80 | | 100 | $-1$ |
| | 30 | 300 | | 225 | $-\frac{1}{2}$ |
| | 31 | 60 | | 75 | $-1$ |
| | | 480 | | 600 | $-1$ |

TABLE I. Parameters for the inverse of the measurement channels for symmetric generalised measurements constructed in Ref. [9] with the same naming convention of the symmetry groups. Notice that here each row corresponds to a single generalised measurement, which is decomposed to several idealised measurements in Ref. [9]. The latter information can be used to simulate the indicated generalised measurement by randomising the corresponding group of idealised measurements.

A large class of uniform and rigidly symmetric measurements beyond qubits are studied in Ref. [9], whose parameters of the classical shadows are enumerated in Table I.

### Appendix F: More detailed on the analysis of noise in the quantum measurements

Readout noise refers to errors that arise due to a misreading of outcomes in the computational basis. Typically, such errors are resulting from decoherence during the measurement process and from overlapping support between the measured physical quantities that correspond to the $|0\rangle$ and $|1\rangle$ state. Errors of that kind have recently attracted a lot of attention and different error mitigation schemes were developed [10–13]. One common model for readout noise is the tensor product noise, where one assumes that the noise acts independently on each qubit. Note that this model does not take into account cross-talk during the readout, i.e., correlated noise between multiple qubits. We define $q_+$ to be the probability that outcome 0 in the computational basis is misread as 1 and $q_-$ to be the probability that outcome 1 in the computational basis is misread as 0. In the case of one

qubit, this can be summarised in a matrix of the form

$$A = \begin{pmatrix} 1 - q_+ & q_- \\ q_+ & 1 - q_- \end{pmatrix} \quad (F1)$$

and $A_{ij}$ is the probability that outcome $j$ is misread as $i$. For a 2-outcome measurement $E = (E_0, E_1)$ intended, the noise $A$ will results in a new actually measurement $\tilde{E}$ with effects of the form $\tilde{E}_0 = A_{00}E_0 + A_{01}E_1$ and $\tilde{E}_1 = A_{10}E_0 + A_{11}E_1$. For the case of randomisation of three Pauli observables considered in the main text, each pair of effects corresponding to the same Pauli observable suffers from this modification due to noise and the effects of the generalised measurement become $1/3\{(1 - q_\pm)|t^\pm\rangle\langle t^\pm| + q_\mp|t^\mp\rangle\langle t^\mp|, t = x, y, z\}$.

In general, the model depends on the chosen implementation, i.e., on the dilation of the measurement. More precisely, often the generalised measurement is implemented by ideal measurement on $n$ ancillary qubits after appropriately coupling to the system. For $n$ ancillary qubits, each can be affected by a different noise and hence the error matrix for the tensor product model is given by $A = A_1 \otimes \cdots \otimes A_n$ where $A_k$ depends on the parameters $q_+^{(k)}$ and $q_-^{(k)}$. The precise effects for the noisy generalised measurement however depend on how precisely the outcomes on the ancillary qubits represent the outcomes of the generalised measurement.

### Appendix G: Optimisation of shadow tomography with simulated annealing

As we described in the main text, the problem of optimising the generalised measurements to minimise the maximal shadow norm $\kappa_E^2(\mathfrak{X})$ for a given set of observables $\mathfrak{X}$ can be carried out with help of simulated annealing. Here we describe how the simulated annealing is implemented for a self-containing reading; for more thorough discussion of the algorithm, we refer the readers to Ref. [14, 15].

The simulated annealing algorithm is best understood as simulating a physical system with states exactly given by coordinates parameterising the effects of the generalised measurements $E = \{E_k\}_{k=1}^N$. The energy of the system is given by the objective function eq. (5) in the main text, namely,

$$\kappa_E^2 = \min_E \{\|X\|_E^2 : X \in \mathfrak{X}\}. \quad (G1)$$

The purpose is then to minimise the energy of the system, solving the optimisation eq. (7) in the main text. Simulated annealing suggests to simulate the system in equilibrium at high temperature, then gradually (quasi-statically) cool the temperature down to zero. Upon the temperature approaching zero, the system converges to the ground state, corresponding to the global minimum of the objective function. Physically, the slow cooling schedule helps the system avoid being stuck in metastable states, which corresponds to local minima [14, 15].

To simulate the system in equilibrium at certain temperature $T$, we use the Hasting-Metropolis algorithm [14]. This is carried out by simulating a large number of elementary steps which mimic the random walk of the system in the parameter space at temperature $T$. At each of the time step, a pair of effects $E_{k_1}$ and $E_{k_2}$ ($1 \leq k_1, k_2 \leq N$, $k_1 \neq k_2$) are chosen at random. An isotropic gaussian random variable in the operator space $\xi$ is drawn. The standard deviation of $\sqrt{T}$ is chosen for $\xi$ to mimic the standard deviation of the brownian motion of a particle at temperature $T$. The state of the system $E$ is modified to $E_{k_1} + \xi$ and $E_{k_2} - \xi$. Notice that this guarantees that the constraint $\sum_{k=1}^N E_k = \mathbb{1}$ is satisfied at any step. The energy at this new state is evaluated and the energy difference $\Delta$ is derived; if either of $E_{k_1} + \xi$ or $E_{k_2} - \xi$ is not positive, an infinite energy difference is assigned. Following the Hasting-Metropolis algorithm, the new state is accepted with certainty if $\Delta \leq 0$, or with probability $\exp(-\Delta/T)$ if $\Delta > 0$; else it is rejected and the system is reversed back to the previous state. After a large number of steps, the system relaxes to equilibrium distribution at temperature $T$ [14]. Then the temperature is cooled down to $0.95T$ and the new equilibrium simulation starts. The procedure is stopped when the temperature of $10^{-8}$ is reached. Randomisation of the starting state and modifications of the parameters are used to guarantee the stability of the obtained optimum. Our codes are available upon reasonable request.

### Appendix H: Proof of the optimality of octahedron measurement for shadow tomography

For a pure state $|\lambda\rangle$, we use $P_\lambda$ to denote the associated projection $P_\lambda = |\lambda\rangle\langle\lambda|$. We consider the problem of predicting all the expectation values of these projections based on classical shadows generated by a measurement $E$. We characterise the predicting power of the shadow tomography by the maximal shadow norm among all these projections,

$$\max_\lambda \|P_\lambda\|_E^2 = \max_\lambda \max_\rho \sum_{k=1}^N \langle\lambda|\hat{\rho}_k|\lambda\rangle^2 \operatorname{Tr}[\rho E_k]. \quad (H1)$$

We are to show that the octahedron measurement is optimal in the sense that it minimises this maximal shadow norm. We assume that the measurement has uniform trace [16]. This implies that $\rho_k$ are all unit trace. In fact, we can derive an explicit formula for the classical shadows in the Bloch representation.

Using the Pauli matrices $\{\sigma_i\}_{i=1}^4 = \{\mathbb{1}, \sigma_x, \sigma_y, \sigma_z\}$ as a basis for the operator space, an operator $X$ on the qubit can be identified with a real vector of 4 components, $X = \frac{1}{2}\sum_{i=1}^4 x_i\sigma_i$, where $x_i = \operatorname{Tr}(\rho\sigma_i)$. Notice that for two operators $X$ and $Y$, which are presented by two vectors $x$ and $y$ respectively, $\operatorname{Tr}(XY) = \frac{1}{2}\sum_{i=1}^4 x_iy_i$. Also, the identity operator $\mathbb{1}$ is identified with the vector $(2, 0, 0, 0)^T$.

In this Bloch representation, each of the effects $E_k$ can be represented by a vector of the form $\frac{2}{N}\begin{pmatrix}1\\\vec{r}_k\end{pmatrix}$, where $\|\vec{r}_k\| \leq 1$ and $\sum_{k=1}^N \vec{r}_k = 0$. Therefore

$$\Phi_E^\dagger = \frac{2}{N}\begin{pmatrix} 1 & 1 & \cdots & 1 \\ \vec{r}_1 & \vec{r}_2 & \cdots & \vec{r}_N \end{pmatrix}. \tag{H2}$$

And the map $\Phi_E$ can be identified with the matrix

$$\Phi_E = \frac{1}{N}\begin{pmatrix} 1 & \vec{r}_1 \\ 1 & \vec{r}_2 \\ \cdot & \cdot \\ 1 & \vec{r}_N \end{pmatrix}. \tag{H3}$$

Explicit calculation then shows that

$$\chi = \begin{pmatrix} 1 & 1 & \cdots & 1 \\ H^{-1}\vec{r}_1 & H^{-1}\vec{r}_2 & \cdots & H^{-1}\vec{r}_N \end{pmatrix}, \tag{H4}$$

where

$$H = \frac{1}{N}\sum_{k=1}^N \vec{r}_k \vec{r}_k^T. \tag{H5}$$

Notice that by definition the columns of $\chi$ are exactly $\hat{\rho}_k$ in the Bloch representation. The explicit formula for the classical shadow allows us to explicitly estimate the shadow norm.

Let us now turn back to the maximal shadow norm (H1). Taking the particular choice of $\rho = |\lambda\rangle\langle\lambda|$ as an ansatz, we have a lower bound

$$\max_\lambda \|P_\lambda\|_E^2 \geq \max_\lambda \sum_{k=1}^N \langle\lambda|\hat{\rho}_k|\lambda\rangle^2 \langle\lambda|E_k|\lambda\rangle. \tag{H6}$$

Then replacing the maximum over $\lambda$ by the average over the Haar measure $\omega$ on the Bloch sphere, one obtains the further lower bound

$$\max_\lambda \|P_\lambda\|_E^2 \geq \int d\omega(\lambda) \sum_{k=1}^N \text{Tr}[(\hat{\rho}_k \otimes \hat{\rho}_k \otimes E_k) P_\lambda^{\otimes 3}]. \tag{H7}$$

It is easy to show that [17]

$$\int d\omega(\lambda) P_\lambda^{\otimes 3} = \frac{1}{12}[(1,2)+(2,3)+(3,1)], \tag{H8}$$

where $(1,2)$, $(2,3)$ and $(3,1)$ denote the operators that permute the corresponding tensor terms in $(\mathbb{C}^2)^{\otimes 3}$. We then arrive at

$$\max_\lambda \|P_\lambda\|_E^2 \geq \frac{1}{12}\sum_{k=1}^N[\text{Tr}(\hat{\rho}_k^2)\text{Tr}(E_k)+2\text{Tr}(\rho_k)\text{Tr}(\hat{\rho}_k E_k)], \tag{H9}$$

where we have used

$$\begin{aligned}\text{Tr}[(\hat{\rho}_k \otimes \hat{\rho}_k \otimes E_k)(1,2)] &= \text{Tr}(\hat{\rho}_k^2)\text{Tr}(E_k),\\ \text{Tr}[(\hat{\rho}_k \otimes \hat{\rho}_k \otimes E_k)(2,3)] &= \text{Tr}(\hat{\rho}_k)\text{Tr}(\hat{\rho}_k E_k),\\ \text{Tr}[(\hat{\rho}_k \otimes \hat{\rho}_k \otimes E_k)(3,1)] &= \text{Tr}(\hat{\rho}_k)\text{Tr}(\hat{\rho}_k E_k).\end{aligned} \tag{H10}$$

Using the explicit Bloch representation for $E_k$ and $\hat{\rho}_k$, we obtain

$$\max_\lambda \|P_\lambda\|_E^2 \geq \frac{1}{12}\left[3 + \frac{1}{N}\sum_{k=1}^N \left(\vec{r}_k^T H^{-2} \vec{r}_k + 2\vec{r}_k^T H^{-1}\vec{r}_k\right)\right]. \tag{H11}$$

Notice that

$$\begin{aligned}\frac{1}{N}\sum_{k=1}^N \vec{r}_k^T H^{-1}\vec{r}_k &= \frac{1}{N}\sum_{k=1}^N \text{Tr}\left(H^{-1}\vec{r}_k\vec{r}_k^T\right)\\ &= \text{Tr}(H^{-1}H)\\ &= 3,\end{aligned}$$

and similarly

$$\frac{1}{N}\sum_{k=1}^N \vec{r}_k^T H^{-2}\vec{r}_k = \text{Tr}(H^{-1}).$$

Then

$$\max_\lambda \|P_\lambda\|_E^2 \geq \frac{1}{12}\left[9 + \text{Tr}(H^{-1})\right]. \tag{H12}$$

Notice that $H$ is positive and denote its positive eigenvalues by $t_1$, $t_2$ and $t_3$, then $\text{Tr}(H^{-1}) = 1/t_1 + 1/t_2 + 1/t_3 \geq 9/(t_1+t_2+t_3) = 9/\text{Tr}(H)$. Moreover, $\text{Tr}(H) = 1/N\sum_{k=1}^N \vec{r}_k^T\vec{r}_k \leq 1$. We then arrive at

$$\max_\lambda \|P_\lambda\|_E^2 \geq \frac{3}{2}. \tag{H13}$$

By direct calculation, one can then show that this inequality is saturated for the octahedron measurement (in which case, $H$ is $1/3$ of the identity matrix). This demonstrates that the octahedron measurement is an optimal choice for the chosen set of observables.

[1] C. M. Bishop, *Pattern Recognition and Machine Learning* (Springer New York, 2006).
[2] M. Guţă, J. Kahn, R. Kueng, and J. A. Tropp, "Fast state tomography with optimal error bounds," J. Phys. A: Math. Theor. **53**, 204001 (2020).
[3] H. Y. Huang, R. Kueng, and J. Preskill, "Predicting many properties of a quantum system from very few measurements," Nat. Phys. **16**, 1050–1057 (2020).

[4] In fact, the convergence of the estimator holds true even if $C_E$ is replaced by any invertible channel $\tilde{C}_E(\rho) = \sum_{k=1}^{N} \text{Tr}(\rho E_k)\tau_k$ for arbitrary operators $\tau_k$. This once more highlights that one should not associate $E_k$ with the state of the system after the measurements.

[5] S. Aaronson, "Shadow tomography of quantum states," SIAM J. Comput. **49**, 368–394 (2020).

[6] A. Acharya, S. Saha, and A. M. Sengupta, "Shadow tomography based on informationally complete positive operator-valued measure," Phys. Rev. A **104**, 052418 (2021).

[7] T. Heinosaari and M. Ziman, *The Mathematical Language of Quantum Theory: From Uncertainty to Entanglement* (Cambridge University Press, 2012).

[8] G. M. D'Ariano, P. Lo Presti, and P. Perinotti, "Classical randomness in quantum measurements," J. Phys. A: Math. Gen. **38**, 5979–5991 (2005).

[9] H. C. Nguyen, S. Designolle, M. Barakat, and O. Gühne, "Symmetries between measurements in quantum mechanics," arXiv:2003.12553 (2020).

[10] S. Bravyi, S. Sheldon, A. Kandala, D. C. Mckay, and J. M. Gambetta, "Mitigating measurement errors in multiqubit experiments," Phys. Rev. A **103**, 042605 (2021).

[11] B. Nachman, M. Urbanek, W. A. de Jong, and C. W. Bauer, "Unfolding quantum computer readout noise," npj Quantum Inf. **6**, 84 (2020).

[12] R. Hicks, B. Kobrin, C. W. Bauer, and B. Nachman, "Active readout-error mitigation," Phys. Rev. A **105**, 012419 (2022).

[13] K. Temme, S. Bravyi, and J. M. Gambetta, "Error mitigation for short-depth quantum circuits," Phys. Rev. Lett. **119**, 180509 (2017).

[14] W. H. Press, S. A. Teukolsky, W. T. Vetterling, and B. P. Flannery, *Numerical recipes* (Cambridge University Press, 2007).

[15] V. Granville, M. Krivanek, and J.-P. Rasson, "Simulated annealing: a proof of convergence," IEEE Trans. Pattern Anal. Mach. Intell. **16**, 652–656 (1994).

[16] This is often not a restrictive assumption, since starting with a measurement with effects of non-uniform traces, by virtually splitting each effect in an appropriate number of identical smaller effects, a measurement with effects with uniform traces can be achieved. Indeed, suppose $E$ is a measurement with difference traces for each effect, $\alpha_k = \text{Tr}(E_k)$. One can always approximate $\alpha_k$ by a rational number. Then one can choose a sufficient small number $\epsilon$ such that $\alpha_k/\epsilon = N_k$ are all integers. Then we split an effect $E_k$ into $N_k$ identical effects $(E_k/N_k, E_k/N_k, \ldots, E_k/N_k)$ and obtain a new measurement of $\sum_{k=1}^{N} N_k$ effects. This measurement is uniform trace by construction.

[17] W. Fulton and J. Harris, *Representation Theory* (Springer New York, 2004).



\end{document}